\documentclass[12pt]{article}
\usepackage{amsmath,multirow}
\usepackage{amssymb}
\usepackage{graphicx}
\usepackage[comma]{natbib}
\usepackage[colorlinks,linkcolor=red,anchorcolor=blue,citecolor=blue]{hyperref}
\usepackage{float}
\usepackage{bm}
\usepackage{longtable}
\usepackage{multirow,booktabs}
\usepackage{lineno,hyperref}
\usepackage{rotating}
\usepackage{chngpage}
\usepackage{color, soul}
\usepackage{diagbox}
\usepackage{lscape}
\usepackage{enumerate}
\usepackage{dsfont}

\newtheorem{theorem}{Theorem}

\newtheorem{remark}{Remark}

\newtheorem{definition}{Definition}
\newtheorem{lemma}{Lemma}
\newtheorem{assumption}{Assumption}
\makeatletter \def\@biblabel#1{#1.} \makeatother
\allowdisplaybreaks

\catcode`@=11 \@addtoreset{equation}{section} \catcode`@=12
\newtheorem{example}{Example}
\begin{document}
\title{Change-point analysis for binomial autoregressive model with application to price stability counts
\footnotetext{$^1$School of Mathematics and Statistics, Liaoning University, Shenyang, China\\
\indent~~$^2$School of Mathematics, Jilin University, Changchun, China\\
\indent~~$^3$School of Mathematics and Statistics, Xi'an Jiaotong University, Xi'an, China\\
\indent~~$\ast$Corresponding author, E-mail: kangyao92@163.com\\
}
}
\author{Danshu Sheng$^1$, Chang Liu$^2$, Yao Kang$^{3*}$}
\date{}
\maketitle
\begin{center}
\begin{minipage}{14.5truecm}
{{\bf Abstract}~~The first-order binomial autoregressive (BAR(1)) model is the most frequently used tool to analyze the bounded count time series.
The BAR(1) model is stationary and assumes process parameters to remain constant throughout the time period, which may be incompatible with the non-stationary real data, which indicates piecewise stationary characteristic.
To better analyze the non-stationary bounded count time series, this article introduces the BAR(1) process with multiple change-points,
which contains the BAR(1) model as a special case.
Our primary goals are not only to detect the change-points, but also to give a solution to estimate the number and locations of the change-points.
For this, the cumulative sum (CUSUM) test and minimum description length (MDL) principle are employed to deal with the testing and estimation problems.
The proposed approaches are also applied to analysis of the
Harmonised Index of Consumer Prices of the European Union.}
\\
~\\
$\mathbf{Keywords:}$ BAR(1) model $\cdot$ Change-point $\cdot$ INAR(1) model $\cdot$ Parameter estimation $\cdot$ CUSUM test
\end{minipage}
\end{center}
\section{Introduction}
In recent years, modeling and analysis of count time series have become an attractive issue with a large quantity of articles in fields like epidemiology, social sciences, economics, life sciences and others.
One of the most commonly used approaches to analyze count time series is to construct integer-valued time series models based on different types of thinning operators.
In particular, the binomial thinning operator is the most popular one in real-world applications since its simplicity and high interpretability.
The binomial thinning operator, which was proposed by \cite{SV1979}, is defined as
\begin{align*}
\alpha\circ X=\sum_{i=1}^{X}B_i,
\end{align*}
where $\alpha\in (0,1)$, $\{B_i\}$ is an independent and identically distributed (i.i.d.) Bernoulli($\alpha$) random sequence independent of non-negative integer-valued random variable $X$.

Based on the different actual background, the research on count time series is mainly divided into statistical inference for unbounded and bounded integer-valued time series model.
On one hand, the unbounded count time series (having a range contained in $\mathbb{N}_{0}=\{0,1,2,...\}$) are frequently encountered in practice, such as monthly unemployment figures, counts of fatal accidents,
severe injury accidents, minor injury accidents, vehicle damage accidents and so on.
First-order integer-valued autoregressive (INAR(1)) model based on the binomial thinning operator is the most commonly applied tool to deal with the unbounded count time series.
We give the definition of the INAR(1) model as follows.
\begin{definition}
The INAR(1) model $\{X_{t}\}_{t\geqslant1}$ is defined by the following recursion
\begin{align*}
X_{t}=\alpha \circ X_{t-1}+\epsilon_{t},~t=1,2,...,
\end{align*}
where $``\circ"$ is the binomial thinning operator,
$\{\epsilon_t\}$ is a sequence of i.i.d. integer-valued random variables and $\epsilon_t$ is not depending on past values of $\{X_{s}\}_{s<t}$.
\end{definition}
Due to the flexibility and practicability of the INAR(1) model, a large quantity of articles focusing on the modeling and statistical inference for the
INAR(1) model have arisen.
For example, \cite{Al1987}, \cite{Scotto2018}, \cite{BS2019}, \cite{Darolles2019}, \cite{Kang2020b,Kang2022,Kang2023} and \cite{Rao2022} considered the modeling of the INAR(1) models to better handle the fitting problems for unbounded count time series.
\cite{Pedeli2015}, \cite{Jentsch2019} studied the parameter estimation for the INAR(1) models.
\cite{McCabe2011}, \cite{Lu2021}, \cite{Freeland2004b}, \cite{Maiti2017} detailedly investigated the prediction approaches for the INAR(1) models.
\cite{SW2014}, \cite{Weiss2019} handled the testing problems for overdispersion and zero inflation in INAR(1) models framework.
\cite{F2016,F2021}, \cite{Henderson2018}, \cite{Guan2022} and \cite{Gourieroux2004} applied the INAR(1) models to under-reported data, longitudinal data and insurance actuarial.

On the other hand, the bounded count time series (with a fixed finite range $\{0,1,2,...,n\}$) are also sometimes suffered, such as
the monitoring of computer pools (with $n$ workstations), infections (with $n$ individuals), metapopulations (with $n$ patches)
and transactions in the stock market (with $n$ listed companies).
The first-order binomial autoregressive (BAR(1)) model, proposed by \cite{McKenzie1985}, is the most natural choice to handle this kind of data.
We give the definition of the BAR(1) model below.
\begin{definition}
The BAR(1) process $\{X_t\}_{t\geqslant1}$ is defined by the recursion
\begin{align}\label{BAR(1)}
X_t=\alpha\circ X_{t-1}+\beta\circ (N-X_{t-1}),~X_0\sim \mathrm{B}(N,p),
\end{align}
where $``\circ"$ is the binomial thinning operator and $N\in\mathbb{N}$ is a known upper bound of the model, $\beta:=p(1-\rho)$, $\alpha:=\beta+\rho$, $p\in(0,1)$, $\rho\in\displaystyle\left(\max{\left\{-\frac{p}{1-p},-\frac{1-p}{p}\right\}},1\right)$.
$\bm{\theta}:=(p,\rho)^{\top}$ is the parameter vector corresponding to the BAR(1) process.
The condition mean and variance of the BAR(1) model are ${\rm E}(X_t|\mathcal{F}_{t-1})=\rho X_{t-1}+Np(1-\rho)$ and ${\rm Var}(X_t|\mathcal{F}_{t-1})=\rho(1-\rho)(1-2p)X_{t-1}+Np(1-\rho)[1-p(1-\rho)]$, respectively,
where $\mathcal{F}_{t-1}$ is the $\sigma$-field generated by the whole information up to time $t-1$.
All thinnings are performed independently of each other, and the thinnings at time $t$ are independent of $\{X_s\}_{s<t}$.
\end{definition}
The BAR(1) model is a strictly stationary and ergodic Markov chain with $h$-step transition probabilities \citep{Weiss2012}
\begin{align}\label{tranp1}
\mathrm{P}(X_t=j|X_{t-h}=i)
 &=\sum_{k=\max\{0,i+j-N\}}^{\min\{i,j\}}\binom{i}{k}\binom{N-i}{j-k}\alpha_{h}^{k}(1-\alpha_{h})^{i-k}\beta_{h}^{j-k}(1-\beta_{h})^{N-i-j+k},
\end{align}
where $\beta_{h}=p(1-\rho^{h})$ and $\alpha_{h}=\beta_{h}+\rho^{h}$.
During the past ten years, the interest in the BAR(1) process has significantly increased and research on this model has gained plentiful and substantial harvest.
For example, \cite{Scotto2014}, \cite{Weiss2014} and \cite{Kang2020b,Kang2021,Kang2023,Kang2024} proposed several extensions to the classical BAR(1) model.
\cite{Weiss2013a,Weiss2013b} studied the parameter estimation for the BAR(1) model.
\cite{Kim2015} and \cite{Kim2018} considered the testing problems for zero inflation and goodness-of-fit in BAR(1) model framework.
\cite{Weiss2012} and \cite{Gouveia2018} applied the BAR(1) model to the ecology, epidemiology and meteorology.

The analysis of change-points, or structural breaks, was initially developed by \cite{Page1954,Page1955} for the detection of change in the mean of independent normal observations.
In the change-points analysis framework, inferential problems primarily involve two aspects, which are respectively detection and estimation.
As for change-points detection, one tests the null hypothesis of no change in the parameters of the statistical model against the alternative hypothesis that parameters of the model change subsequent to at least one unknown change-points.
With regard to change-point estimation, researchers are not only interested in the number of change-points, but also attach importance to obtain the location of the change-points.
As the continuous improvement of relevant theories, researchers have come to realize that it appears to be of significant importance to incorporate dependent observations into the change-points analysis since the non-stationary time series, which indicated piecewise stationary characteristic, were frequently encountered.
During the past few decades, the time series change-points analysis has been vigorously developed and we refer to the studies of \cite{Aue2013}, \cite{Jandhyala2013} and \cite{Aminikhanghahi2017} for a general review.
Especially, relevant achievements in count time series gradually become abundant and
articles concentrating on the change-points analysis for the INAR(1) models frequently arise in recent years.
For example,
\cite{Pap2013} proposed sever test statistics to detect the change-points in the INAR(1) model.
\cite{Kang2009}, \cite{YuandKim2020} and \cite{Lee2022} considered the problem of testing for a parameter change in different types of INAR(1) models by taking advantage of the CUSUM test.
\cite{Kashikar2013}	developed the Poisson INAR($p$) process with change-points and applied it to the two biometrical data sets.
\cite{Chattopadhyay2021} considered the problem of change-point analysis for the INAR(1) model with time-varying covariates.
\cite{Yu2022} applied the empirical likelihood ratio (ELR) test to uncover a structural change in INAR processes,
\cite{Sheng2023} studied the change-points analysis of the MCP-GCINAR model based on the MDL principle, and optimized by genetic algorithm (GA).
For a review of the change-points analysis in count time series, we refer to the survey by \cite{Lee2021}.

However, to our best knowledge, the change-points analysis for time series of counts is mainly considered in unbounded count data.
The related research is rare in bounded count data context, although the development would be extremely important for practice.
Up to now, there is only one article that studied the relevant issue.
\cite{Zhang2023} proposed a BAR(1) model with one change-point and
further studied the change-point detection and estimation problems.
However, it is well-known that the multiple change-points model is an extension to the one change-point model. Moreover, a significant limitation of one change-point model is that the multiple change-points situation is more commonly observed in practice.
Based on the above consideration, we conclude that it is a vitally necessary and significant issue to come up with a solution to modeling the bounded count time series with multiple change-points.
To further illustrate the above statement, we consider the following example.
\begin{example}
\cite{WeissKim2014} studied the data set representing how many of the seventeen European Union countries have a monthly inflation rate below 2$\%$ from January 2000 to December 2011. The data set has a range in $\{0,1,2,...,17\}$ and the length of data is 132. Figure \ref{F1} shows the time series plot, autocorrelation function (ACF), and partial ACF (PACF).
As pointed out by \cite{WeissKim2014}, they only analyzed the first 84 observations, i.e., the data corresponds to the period 2000-2006,
since these data give no reason to doubt a stationarity based on the time series plot.
The authors also agreed with the opinion that the stationarity of the series is violated due to the occurrence of economic events since 2007 such as sub-prime crisis and so on.
Moreover, based on the sample path in Figure \ref{F1}, it is not difficult to speculate that a bounded count time series model with multiple change-points is a reasonable choice to analyze this data set.
\end{example}
To achieve the goal of better fitting the non-stationary bounded count time series,
this article concentrates on the statistical analysis for the BAR(1) model with multiple change-points.
For this, we initially give the definition of the BAR(1) model with multiple change-points and study the statistical inference for the proposed model.
Furthermore, the CUSUM test based on the conditional least squares and modified quasi-likelihood estimators are used to detect the change-points.
The estimation problems for the number and location of change-points are also handled by utilizing the minimum description length principle.
The number and location of change-points are implicitly defined as the optimizer of an objective function, and the searching algorithm based on genetic algorithm is proposed to solve this difficult optimization problem.

The rest contents of this article are organized as follows.
In Section 2, the CUSUM test based on the conditional least squares and modified quasi-likelihood estimators is employed to detect the change-points in the BAR(1) model.
In Section 3, the definition of the BAR(1) model with multiple change-points is proposed.
Furthermore, we utilize the minimum description length criterion to determine the number and locations of the change-points.
Section 4 evaluates the proposed test statistics and estimation methods via some simulation studies.
In Section 5, to show the usefulness of our model and methods, we apply the proposed model to the monthly price stability counts.
The article ends with a conclusion section and all proofs are given in Appendix.

\section{CUSUM test for change-point detection}\label{sect2}
In this section, we focus on the change-point detection in the BAR(1) model.
The CUSUM test based on conditional least squares (CLS), modified quasi-likelihood (MQL), and conditional maximum likelihood (CML) estimators
are employed.
To task it, we set up the null and alternative hypotheses as follows:
$$\mbox{$\mathcal{H}_0$: $\rho$ and $p$  do not change over $X_1,...,X_{n}$ v.s. $\mathcal{H}_1$: not $\mathcal{H}_0$.}$$
Under $\mathcal{H}_0$, we denote the parameter vector by $\bm{\theta}=(\rho,p)^{\top}$,
the parameter space by
$$\Theta_{\text{c}}=[\delta,1-\delta]\times\displaystyle\left[\max{\left\{-\frac{p}{1-p},-\frac{1-p}{p}\right\}}+\delta,1-\delta\right],$$
where $\delta$ is a finite positive constant.
\subsection{CUSUM test based on CLS estimators}
Suppose we have a series of observations $\{X_t\}_{t=1}^n$ generated from the BAR(1) process under $\mathcal{H}_0$.
The condition mean of the BAR(1) model is ${\rm E}(X_t|\mathcal{F}_{t-1})=\rho X_{t-1}+Np(1-\rho)$.
Then, the CLS estimator $\hat{\bm{\theta}}_{n,CLS}:=(\hat{\rho}_{n,CLS},\hat{p}_{n,CLS})^{\top}$ of $\bm{\theta}$ is obtained by minimizing the sum of the squared deviations
\begin{align}\label{Sn_cls}
S_{n}(\bm{\theta})&:=\sum_{t=1}^{n}[X_t-\rho X_{t-1}-Np(1-\rho)]^2=\sum_{t=1}^{n}s_{t}^{2}(\bm{\theta}),
\end{align}
The closed-form expressions for the CLS estimators can be given by
\begin{equation}\label{cls_closed-form}
\begin{split}
\hat{p}_{n,CLS}&=\frac{\sum_{t=1}^nX_t-\hat{\rho}_{n,CLS}\sum_{t=1}^nX_{t-1}}{nN(1-\hat{\rho}_{n,CLS})},\\
\hat{\rho}_{n,CLS}&=\frac{n\sum_{t=1}^nX_tX_{t-1}-\left(\sum_{t=1}^nX_t\right)\left(\sum_{t=1}^nX_{t-1}\right)}{n\sum_{t=1}^nX_{t-1}^2-\left(\sum_{t=1}^nX_{t-1}\right)^2}.
\end{split}
\end{equation}
Since the BAR(1) model is stationary, ergodic, and all moments are bounded,
then using the Taylor expansion and the martingale central limit theorem, the following theorem about the consistency and asymptotic normality of the parameter estimator $\hat{\bm{\theta}}_{n,CLS}$ can be obtained. The detailed proof is presented in the Appendix.
\begin{theorem}\label{clsnormal}
Let $\bm{\theta}_{0}=(\rho_{0},p_{0})^{\top}$ be the true value of the parameter vector $\bm{\theta}$. Suppose that $\bm{\theta}_0$ is an interior point of the compact space, then the CLS estimator $\hat{\bm{\theta}}_{n,CLS}$ satisfies the following asymptotic normality
\begin{align*}
\sqrt{n}\big(\hat{\bm{\theta}}_{n,CLS}-\bm{\theta}_{0}\big)\mathop{\longrightarrow}\limits^{\mathrm{d}} \mathrm{N}\left(\bm{0},\bm{V}_{CLS}^{-1}\bm{W}_{CLS}\bm{V}_{CLS}^{-1}\right),
\end{align*}
as $n\rightarrow\infty$, where $\bm{W}_{CLS}$ and $\bm{V}_{CLS}$ have the entries $W_{CLS}^{(i,j)}$ and $V_{CLS}^{(i,j)}$, $i,j=1,2$, with
\begin{align*}
W_{CLS}^{(1,1)}&=\mathrm{E}[s_{t}(\bm{\theta}_{0})(X_{t-1}-Np_{0})]^2,~
W_{CLS}^{(2,2)}=\mathrm{E}[s_{t}(\bm{\theta}_{0})N(1-\rho_{0})]^2,\\
W_{CLS}^{(1,2)}&=W_{CLS}^{(2,1)}=\mathrm{E}\{[N(1-\rho_{0})(X_{t-1}-Np_{0})]s_{t}^{2}(\bm{\theta}_{0})\},\\
V_{CLS}^{(1,1)}&=-\mathrm{E}(X_{t-1}-Np_{0})^2,~
V_{CLS}^{(1,2)}=V_{CLS}^{(2,1)}=0,~
V_{CLS}^{(2,2)}=-[N(1-\rho_{0})]^{2}.
\end{align*}
\end{theorem}
Motivated by \cite{Kang2009} and \cite{Lee2016}, applying the relationship between Brownian motion and Brownian bridge, we obtain the following conclusion.
\begin{theorem}\label{test1}
Under $\mathcal{H}_0$,
$$\bm{W}_{CLS}^{-1/2}\bm{V}_{CLS}\frac{[n\lambda]}{\sqrt{n}}(\hat{\bm{\theta}}_{[n\lambda],CLS}-\hat{\bm{\theta}}_{n,CLS}) \mathop{\longrightarrow}\limits^{\mathrm{d}}\bm{B}_2(\lambda),$$
as $n\rightarrow\infty$, where $[n\lambda]$ is the greatest integer that is less than or equal to $n\lambda$ and $0\leqslant \lambda\leqslant1$,
$\bm{B}_2(\lambda)=(B_1(\lambda),B_2(\lambda))^{\top}$ is a two-dimensional Brownian bridge.
\end{theorem}
In fact, according to the ergodicity of the BAR(1) model, it is easy to see that
\begin{align*}
\hat{\bm{V}}_{n,CLS}=\left( {\begin{array}{lc}
{\frac{1}{n}\sum_{t=1}^n(X_{t-1}-N\hat{p}_{n,CLS})^2}&{\frac{1}{n}\sum_{t=1}^{n}(X_{t-1}-N\hat{p}_{n,CLS})N(1-\hat{\rho}_{n,CLS})}\\
{\frac{1}{n}\sum_{t=1}^n(X_{t-1}-N\hat{p}_{n,CLS})N(1-\hat{\rho}_{n,CLS})}&{N^2(1-\hat{\rho}_{n,CLS})^2}
\end{array}}\right)
\end{align*}
is a consistent estimator of $\bm{V}_{CLS}$, and
\begin{align*}
\hat{\bm{W}}_{n,CLS}=\left({\begin{array}{*{2}{c}}
\hat{{W}}_{n,CLS}^{(1,1)}&\hat{{W}}_{n,CLS}^{(1,2)}\\
\hat{{W}}_{n,CLS}^{(2,1)}&\hat{{W}}_{n,CLS}^{(2,2)}
\end{array}}\right),
\end{align*}
where
\begin{align*}
\hat{{W}}_{n,CLS}^{(1,1)}&={\frac{1}{n}\sum_{t=1}^{n}[s_t(\hat{\bm{\theta}}_{n,CLS})(X_{t-1}-N\hat{p}_{n,CLS})]^2},\\
\hat{{W}}_{n,CLS}^{(1,2)}&=\hat{W}_{n,CLS}^{(2,1)}={\frac{1}{n}\sum_{t=1}^{n}s_t^2(\hat{\bm{\theta}}_{n,CLS})(X_{t-1}-N\hat{p}_{n,CLS})N(1-\hat{\rho}_{n,CLS})},\\
\hat{{W}}_{n,CLS}^{(2,2)}&={\frac{1}{n}\sum_{t=1}^{n}[s_t(\hat{\bm{\theta}}_{n,CLS})N(1-\hat{\rho}_{n,CLS})]^2},
\end{align*}
is a consistent estimator of $\bm{W}_{CLS}$.
Thus, we have the following result under $\mathcal{H}_0$:
\begin{align*}
\hat{\bm{W}}_{n,CLS}^{-1/2}\hat{\bm{V}}_{n,CLS}\frac{[n\lambda]}{\sqrt{n}}(\hat{\bm{\theta}}_{[n\lambda],CLS}-\hat{\bm{\theta}}_{n,CLS})
\mathop{\longrightarrow}\limits^{\mathrm{d}}\bm{B}_2(\lambda),
\end{align*}
as $n\rightarrow\infty$. Thus, we can construct the CUSUM test based on the CLS estimators and deduce its asymptotic distribution.
\begin{theorem}\label{the_cls_c}
Let $k_0$ be a positive integer, and define
\begin{align*}
C_{n}^{CLS}=\max_{k_0\leqslant k\leqslant n}\frac{k^2}{n}(\hat{\bm{\theta}}_{k,CLS}-\hat{\bm{\theta}}_{n,CLS})^{\top}\hat{\bm{V}}_{n,CLS}\hat{\bm{W}}_{n,CLS}^{-1}\hat{\bm{V}}_{n,CLS}(\hat{\bm{\theta}}_{k,CLS}-\hat{\bm{\theta}}_{n,CLS}).
\end{align*}
Then, under $\mathcal{H}_0$,
$$C_n^{CLS}\mathop{\longrightarrow}\limits^{\mathrm{d}}\sup_{0\leqslant \lambda\leqslant1}||\bm{B}_2(\lambda)||^2,~~n\rightarrow\infty.$$
Under $\mathcal{H}_1$,
$$C_n^{CLS}\mathop{\longrightarrow}\limits^{\mathrm{p}}+\infty,~~n\rightarrow\infty.$$
\end{theorem}
\subsection{CUSUM test based on MQL estimators}
Similar to the CUSUM test based on the CLS estimators, we can also construct the CUSUM test from the asymptotic distribution of the MQL estimators.
Let
\begin{align*}
&D_{t}^{-1}(\hat{\bm{\theta}}_{n,CLS}):=\mathrm{Var}(X_{t}|\mathcal{F}_{t-1})\\
&=\hat\rho_{n,CLS}(1-\hat\rho_{n,CLS})(1-2\hat p_{n,CLS})X_{t-1}+N\hat p_{n,CLS}(1-\hat\rho_{n,CLS})[1-\hat p_{n,CLS}(1-\hat\rho_{n,CLS})].
\end{align*}
Then the MQL estimator $\hat{\bm{\theta}}_{n,MQL}:=(\hat{\rho}_{n,MQL},\hat{p}_{n,MQL})^{\top}$ of $\bm{\theta}$ is obtained by minimizing the sum of the squared deviations
\begin{align*}
Q_{n}(\bm{\theta})&:=\sum_{t=1}^{n}D_{t}(\hat{\bm{\theta}}_{n,CLS})[X_t-\rho X_{t-1}-Np(1-\rho)]^2=\sum_{t=1}^{n}D_{t}(\hat{\bm{\theta}}_{n,CLS})s_{t}^2(\bm{\theta}).
\end{align*}
The closed-form expressions for the MQL estimator can be given by
\begin{align*}
\hat{\rho}_{n,MQL}&=\frac{\sum\limits_{t=1}^{n}D_{t}(\hat{\bm{\theta}}_{n,CLS})\sum\limits_{t=1}^{n}D_{t}(\hat{\bm{\theta}}_{n,CLS})X_{t-1}X_{t}-\left(\sum\limits_{t=1}^{n}D_{t}(\hat{\bm{\theta}}_{n,CLS})X_{t}\right)\left(\sum\limits_{t=1}^{n}D_{t}(\hat{\bm{\theta}}_{n,CLS})X_{t-1}\right)}
{\sum\limits_{t=1}^{n}D_{t}(\hat{\bm{\theta}}_{n,CLS})\sum\limits_{t=1}^{n}D_{t}(\hat{\bm{\theta}}_{n,CLS})X_{t-1}^{2}-\left(\sum\limits_{t=1}^{n}D_{t}(\hat{\bm{\theta}}_{n,CLS})X_{t-1}\right)^2},\\
\hat{p}_{n,MQL}&=\frac{\sum\limits_{t=1}^nD_{t}(\hat{\bm{\theta}}_{n,CLS})X_t-\hat{\rho}_{n,MQL}\sum\limits_{t=1}^nD_{t}(\hat{\bm{\theta}}_{n,CLS})X_{t-1}}{N(1-\hat{\rho}_{n,MQL})\sum\limits_{t=1}^nD_{t}(\hat{\bm{\theta}}_{n,CLS})},
\end{align*}
Similar to the proof of Theorem \ref{clsnormal}, $\hat{\bm{\theta}}_{n,MQL}$ is a consistent estimator and satisfies asymptotic normality.
\begin{theorem}
Suppose $\bm{\theta}_{0}=(\rho_{0},p_{0})^{\top}$ is the true value of the parameter vector $\bm{\theta}$, 
then the MQL estimator $\hat{\bm{\theta}}_{n,MQL}$ satisfies the following asymptotic normality
\begin{align*}
\sqrt{n}\big(\hat{\bm{\theta}}_{n,MQL}-\bm{\theta}_{0}\big)\mathop{\longrightarrow}\limits^{\mathrm{d}} \mathrm{N}\left(\bm{0},\bm{V}_{MQL}^{-1}\bm{W}_{MQL}\bm{V}_{MQL}^{-1}\right),
\end{align*}
as $n\rightarrow\infty$, where $\bm{W}_{CLS}$ and $\bm{V}_{CLS}$ have the entries $W_{MQL}^{(i,j)}$ and $V_{MQL}^{(i,j)}$, $i,j=1,2$, with
\begin{align*}
W_{MQL}^{(1,1)}&=\mathrm{E}[D_{t}(\hat{\bm{\theta}}_{n,CLS})s_{t}(\bm{\theta}_{0})(X_{t-1}-Np_{0})]^2,~
W_{MQL}^{(2,2)}=\mathrm{E}[D_{t}(\hat{\bm{\theta}}_{n,CLS})s_{t}(\bm{\theta}_{0})N(1-\rho_{0})]^2,\\
W_{MQL}^{(1,2)}&=W_{MQL}^{(2,1)}=\mathrm{E}\{D_{t}^{2}(\hat{\bm{\theta}}_{n,CLS})[N(1-\rho_{0})(X_{t-1}-Np_{0})]s_{t}^{2}(\bm{\theta}_{0})\},\\
V_{MQL}^{(1,1)}&=-\mathrm{E}[D_{t}(\hat{\bm{\theta}}_{n,CLS})(X_{t-1}-Np_{0})^2],~
V_{MQL}^{(2,2)}=-[N(1-\rho_{0})]^{2}\mathrm{E}[D_{t}(\hat{\bm{\theta}}_{n,CLS})],\\
V_{MQL}^{(1,2)}&=V_{MQL}^{(2,1)}=-\mathrm{E}[D_{t}(\hat{\bm{\theta}}_{n,CLS})N(1-\rho_{0})(X_{t-1}-Np_{0})].
\end{align*}
\end{theorem}
Similar, we can also approximate ${\bm{V}}_{MQL}$ and ${\bm{W}}_{MQL}$ by their consistent estimators $\hat{\bm{V}}_{n,MQL}$ and $\hat{\bm{W}}_{n,MQL}$,
and then obtain the following result.
\begin{theorem}\label{the_qml_c}
Let $k_0$ be a positive integer, and define
\begin{align*}
C_{n}^{MQL}=\max_{k_0\leqslant k\leqslant n}\frac{k^2}{n}(\hat{\bm{\theta}}_{k,MQL}-\hat{\bm{\theta}}_{n,MQL})^{\top}\hat{\bm{V}}_{n,MQL}\hat{\bm{W}}_{n,MQL}^{-1}\hat{\bm{V}}_{n,MQL}(\hat{\bm{\theta}}_{k,MQL}-\hat{\bm{\theta}}_{n,MQL}),
\end{align*}
where the expressions of $\hat{\bm{V}}_{n,MQL}$ and $\hat{\bm{W}}_{n,MQL}$ are in Appendix. Then, under $\mathcal{H}_0$,
$$C_n^{MQL}\mathop{\longrightarrow}\limits^{\mathrm{d}}\sup_{0\leqslant s\leqslant1}||\bm{B}_2(\lambda)||^2,~~n\rightarrow\infty.$$
Under $\mathcal{H}_1$,
$$C_n^{MQL}\mathop{\longrightarrow}\limits^{\mathrm{p}}+\infty,~~n\rightarrow\infty.$$
\end{theorem}
\begin{remark}
The MQL estimation method can be considered a form of weighted CLS method.
Clearly, it exhibits lower asymptotic variance relative to the CLS method.
This suggests greater efficiency in estimating parameters with the MQL method.
Hence, such efficiency gains may enhance the power of CUSUM tests utilizing MQL estimators, as supported by subsequent simulation studies.
\end{remark}

\subsection{CUSUM test based on CML estimators}
In this subsection, we review the conditional maximum likelihood (CML) estimation for the BAR(1) model.
The log-likelihood function for $\{X_{t}\}_{t=i+1}^{n}$ can be given by
\begin{align*}
L_{n}(i,\bm{\theta})&=\sum\limits_{t=i+1}^{n}\ell_t(\bm{\theta}|X_{s},s<t)=\sum\limits_{t=i+1}^{n}\log\mathrm{P}(X_{t}|X_{t-1}),
\end{align*}
where $\mathrm{P}(X_{t}|X_{t-1})$ is defined in (\ref{tranp1}) with $h=1$. 
The CML estimator $\hat{\bm{\theta}}_{n,CML}:=(\hat{\rho}_{n,CML},\hat{p}_{n,CML})^{\top}$ of $\bm{\theta}$ for $\{X_{t}\}_{t=1}^{n}$ is obtained by
\begin{align*}
\hat{\bm{\theta}}_{n,CML}=\arg\max\limits_{\bm{\theta}\in\Theta_c} L_{n}(0,\bm{\theta}).
\end{align*}
According to the discussion in Section 2 of \cite{Weiss2013b}, $\hat{\bm{\theta}}_{n,CML}$ is consistent and has the following asymptotically distribution,
\begin{align*}
\sqrt{n}(\hat{\bm{\theta}}_{n,CML}-\bm{\theta}_0)\mathop{\longrightarrow}\limits^{\mathrm{d}} N(0,\bm{I}^{-1}(\bm{\theta}_0)),~n\rightarrow \infty,
\end{align*}
where $\bm{\theta}_0$ denotes the true parameter value of $\bm{\theta}$ and
$\bm{I}^{-1}(\bm{\theta}_0)=\mathrm{E}\left[-\frac{\partial^2 \ell_t(\bm{\theta}_0|X_{t-1})}{\partial \bm{\theta}\partial \bm{\theta}^{\mathrm{T}}}\right]$ denotes the Fisher information matrix.
Let $\hat{\bm{I}}_{n,CML}$ be the approximation of $\bm{I}(\bm{\theta}_0)$.
Then, analogy to the process in the previous two sections, we obtain the following Theorem.
\begin{theorem}\label{the_cml_c}
Let $k_0$ be a positive integer, and define
\begin{align*}
C_{n}^{CML}=\max_{k_0\leqslant k\leqslant n}\frac{k^2}{n}(\hat{\bm{\theta}}_{k,CML}-\hat{\bm{\theta}}_{n,CML})^{\top}\hat{\bm{I}}_{n,CML}(\hat{\bm{\theta}}_{k,CML}-\hat{\bm{\theta}}_{n,CML}).
\end{align*}
where the expressions of $\hat{\bm{I}}_{n,CML}$ is in Appendix. Then, under $\mathcal{H}_0$,
$$C_n^{CML}\mathop{\longrightarrow}\limits^{\mathrm{d}}\sup_{0\leqslant s\leqslant1}||\bm{B}_2(\lambda)||^2,~~n\rightarrow\infty.$$
Under $\mathcal{H}_1$,
$$C_n^{CML}\mathop{\longrightarrow}\limits^{\mathrm{p}}+\infty,~~n\rightarrow\infty.$$
\end{theorem}
\section{Estimation for the change-points}
Section \ref{sect2} gives a solution to the problem of change-point detection.
However, in many practical applications, estimation for the number and location of change-points is also an important topic.
A reliable estimation method for the change-points will help us in model fitting and analyzing the background of the change-points.
Clearly, if there is only one change-point, applying the one-by-one search method can solve the estimation problem \citep{Zhang2023}.
However, the situation of multiple change-points is commonly encountered in data analysis.
In this case, the number and location of change-points are all unknown and the one-by-one search method will be very inefficient since the high computing cost and the increase in estimation inaccuracy are not negligible.
Thus, in this section, we focus on the estimation for the number and location of change-points.
The minimum description length criterion is used to handle the above concerned issue and a new algorithm, named searching algorithm based on genetic algorithm (S-GA), is proposed.

To better fit the non-stationary bounded count time series, we extend the BAR(1) model defined in Equation (\ref{BAR(1)}) to
the BAR(1) model with $m$ change-points by allowing the parameters of the process to vary according to time.
\begin{definition}
The multiple change-points BAR(1) (MCP-BAR(1)) process with $m$ change-points $\{X_t\}_{t=1}^{n}$ is defined by the recursion:
\begin{align}\label{BAR}
X_{t}
=\begin{cases}
\alpha_{1}\circ X_{t-\tau_{0}-1,1}+\beta_{1}\circ (N-X_{t-\tau_{0}-1,1}),~t\leqslant\tau_{1},\\
\alpha_{2}\circ X_{t-\tau_{1}-1,2}+\beta_{2}\circ (N-X_{t-\tau_{1}-1,2}),~\tau_{1}<t\leqslant \tau_{2},\\
\vdots\\
\alpha_{m+1}\circ X_{t-\tau_{m}-1,m+1}+\beta_{m+1}\circ (N-X_{t-\tau_{m}-1,m+1}),~t>\tau_{m},\\
\end{cases}
\end{align}
where $N\in\mathbb{N}$ is a known upper bound of the model, $\beta_{j}:=p_{j}(1-\rho_{j})$, $\alpha_{j}:=\beta_{j}+\rho_{j}$, $p_{j}\in(0,1)$,  $\rho_{j}\in\displaystyle\left(\max{\left\{-\frac{p_{j}}{1-p_{j}},-\frac{1-p_{j}}{p_{j}}\right\}},1\right)$,
$\bm{\theta}_j:=(p_j,\rho_j)^{\top}$ is the parameter vector corresponding to the $j$th segment of the BAR(1) process
and $\min_{1\leqslant j\leqslant m}||\bm{\theta}_{j+1}- \bm{\theta}_{j}||>0$,
$\bm{\tau}=(\tau_1,\tau_2,...,\tau_{m})^{\top}$ denotes the vector of unknown locations of change-points, $\tau_0=0$ and $\tau_{m+1}=n$.
Each change-points location $\tau_j$, $j=1,...,m$, is an integer between $1$ and $n-1$ inclusive, and the change-points are ordered such that $\tau_{j_1}< \tau_{j_2}$ if, and only if, $j_1<j_2$.
All thinnings are performed independently of each other, and the thinnings at time $t$ are independent of $\{X_s\}_{s<t}$.
\end{definition}

For a specified vector $(m,\bm{\tau})$, the time series $\bm{X}_n=(X_1, X_2,..., X_n)$ generated by (\ref{BAR}) can also be written as
\begin{align}\label{xn}
\bm{X}_n=({\bm X}_{n_1,1}^{\top},...,{\bm X}_{n_j,j}^{\top},...,{\bm X}_{n_{m+1},m+1}^{\top}),
\end{align}
where ${\bm X}_{n_j,j}=(X_{1,j},...,X_{n_j,j})^{\top}$ denotes the time series for the $j$th segment, corresponding to the period $\tau_{j-1}+1\leqslant t\leqslant \tau_{j}$, $n_j=\tau_j-\tau_{j-1}$ for $j= 1,...,m+1$ and $n = n_1 + n_2 +\cdots+ n_{m+1}$. $\{X_{t}\}_{t=1}^{n}$ is a non-stationary process in general but can be viewed as a stationary BAR(1) process in each regime.
%

\subsection{Minimum description length criterion}\label{Set3}
Loosely speaking, change-point estimation can be considered as the identification of points within a data set where the statistical properties change.
One common approach is to minimize a specific information criterion (IC) to identify multiple change-points.
In this section, we apply the minimum description length (MDL) principle of \cite{Rissanen1989} as IC to identify a best-fitting model.

For the sake of readability, we provide the following notations and assumptions, and their corresponding explanations before introducing the MDL criterion.
\\
~\\
$\bm{\mathrm{Notations:}}$
\begin{itemize}
\setlength{\itemsep}{0pt}
\setlength{\parskip}{0pt} 
\setlength{\parsep}{0pt} 
\item Denote this whole class of the MCP-BAR(1) models by $\mathcal{M}$ and any submodel from this class by $\mathcal{F} \in\mathcal{M}$.
\item Let ${\bm\lambda}=(\lambda_1,...,\lambda_{m})$, $0<\lambda_1<\cdots<\lambda_{m}<1$, satisfy $\tau_j=[\lambda_jn]$, where $[x]$ is the greatest integer that is less than or equal to $x$.
\item Denote the true number of change-points by $m_0$, the true location of change-points by $\bm{\lambda}^0=(\lambda_1^0,...,\lambda_{m_0}^0)^{\top}$,
the true parameter vector by ${\bm\theta}^0=\left((\bm{\theta}_1^0)^{\top},...,(\bm{\theta}_{m_0+1}^0)^{\top}\right)^{\top}$ with $\bm{\theta}_j^0=(p_j^0,\rho_j^0)^{\top}$, and $\min_{1\leqslant j\leqslant m_0}||\bm{\theta}_{j+1}^0- \bm{\theta}_{j}^0||>0$.
\end{itemize}

\begin{assumption}\label{assumption1}
To accurately estimate the specified BAR(1) parameter values, the segments must have a sufficient number of observations,
if not, the estimation is overdetermined and the likelihood has an infinite value. So to ensure identifiability of the change-points, when we search for the change-points, we assume that there exists a $\epsilon_{\lambda}>0$ such that $\epsilon_{\lambda}<\min_{1\leqslant j\leqslant m}(|\lambda_j-\lambda_{j-1}|)$ and set
\begin{align*}
A_{\epsilon_{\lambda}}^m=\{{\bm\lambda}\in(0,1)^{m},0<\lambda_1<\cdots<\lambda_{m}<1,\lambda_{j}-\lambda_{j-1}\geqslant
\epsilon_{\lambda},j=1,...,m+1\}.
\end{align*}
so under this restriction the number of change points is bounded by $M_0=[1/\epsilon_{\lambda}]+1$.
\end{assumption}
\begin{assumption}\label{assumption2}
Denote the parameter vector of the $j$th segment by
$\bm{\theta}_j=(p_j,\rho_j)^{\top}$,
which is assumed to be an interior point of the compact space $\Theta_j$,
\begin{align*}
\Theta_j=[\delta,1-\delta]\times\displaystyle\left[\max{\left\{-\frac{p_{j}}{1-p_{j}},-\frac{1-p_{j}}{p_{j}}\right\}}+\delta,1-\delta\right],
\end{align*}
where $\delta$ is a finite positive constant.
$\bm{\theta}=(\bm{\theta}_1^{\top},...,\bm{\theta}_{m+1}^{\top})^{\top}$ belongs to the parameter space $\Theta=\prod\limits_{j=1}^{m+1}\Theta_j$.
\end{assumption}
\begin{assumption}\label{assumption3}
To make sure the change-points exist,
assume that there exists a $\epsilon_{\theta}>0$ such that $\min_{1\leqslant j\leqslant m_0}||\bm{\theta}_{j+1}- \bm{\theta}_{j}||>\epsilon_{\theta}$.
\end{assumption}

Next, we introduce the MDL criterion for the MCP-BAR(1) model.
According to \cite{Davis2006}, the MDL criterion can be regarded as a cost function ($CF$), which is the sum of negative log-likelihood for each of the segments, plus a penalty term. That is, denote a fitted model by $\hat{\mathcal{F}}$,
$$CF_{\hat{\mathcal{F}}}(\bm{X}_n)=-\sum\limits_{j=1}^{m+1}L_{n_j}(\tau_{j-1},\bm{\theta}_j)+CL_{\hat{\mathcal{F}}}(\hat{\mathcal{F}}),$$
where $CL_{\hat{\mathcal{F}}}(\hat{\mathcal{F}})$ is the code length of the fitted model $\hat{\mathcal{F}}$, or called it the penalty term of fitted model $\hat{\mathcal{F}}$.
Next, the task is to derive expressions for $CL_{\hat{\mathcal{F}}}(\hat{\mathcal{F}})$ according to the MDL principle.
Since $\hat{\mathcal{F}}$ is composed of $m$, $\tau_j$'s, $\bm{\theta}_j$'s, we can further decompose $CL_{\hat{\mathcal{F}}}(\hat{\mathcal{F}})$ into
\begin{align*}
CL_{\hat{\mathcal{F}}}(\hat{\mathcal{F}})=&CL_{\hat{\mathcal{F}}}(m)+CL_{\hat{\mathcal{F}}}(\bm{\tau})+CL_{\hat{\mathcal{F}}}(\hat{\bm{\theta}}_1)+\cdots+CL_{\hat{\mathcal{F}}}(\hat{\bm{\theta}}_j)+CL_{\hat{\mathcal{F}}}(\hat{\bm{\theta}}_{m+1}),
\end{align*}
where the first two items are $CL_{\hat{\mathcal{F}}}(m)=\log(m)$ and $CL_{\hat{\mathcal{F}}}(\bm{\tau})=(m+1)\log(n)$.
To calculate $CL_{\hat{\mathcal{F}}}(\hat{\bm{\theta}}_j)$, we use the result of \cite{Rissanen1989}: A maximum likelihood estimator of a real parameter computed from $N$ observations can be effectively encoded with $1/2\log(N)$ bits.
Because each of the two parameters of $\hat{\bm{\theta}}_j$ is computed from $n_j$ observations, there is
$CL_{\hat{\mathcal{F}}}(\hat{\bm{\theta}}_j)=2/2\log(n_j)$.
Then, combining these results, the MDL criterion for the MCP-BAR(1) model is given by
\begin{align}\label{MDL}
\textbf{MDL}(m,\bm{\lambda}, \bm{\theta})=&\log(m)+(m+1)\log n+\sum\limits_{j=1}^{m+1}\log(n_j)-\sum\limits_{j=1}^{m+1}L_{n_j}(\tau_{j-1},\bm{\theta}_j).
\end{align}
The estimator of the number of change-points, the locations of change-points and the parameters in each of the segments, ($\hat{m}_n,\hat{\bm{\lambda}}_n, \hat{\bm{\theta}}_n$), is obtained by
\begin{align}\label{MDL_est}
(\hat{m}_n,\hat{\bm{\lambda}}_n,\hat{\bm{\theta}}_n)=\arg\min\limits_{
m\leq M_0,\bm{\lambda}\in A_{\epsilon_{\lambda}}^m, \bm{\theta}\in\Theta}
 \textbf{MDL}(m,\bm{\lambda},\bm{\theta}).
\end{align}
where $\hat{\bm{\lambda}}_n=(\hat{\lambda}_1,...,\hat{\lambda}_{\hat{m}_n}),~
\hat{\bm{\theta}}_n=(\hat{\bm{\theta}}_1,...,\hat{\bm{\theta}}_j,...,\hat{\bm{\theta}}_{\hat{m}_n+1})$ with
$\hat{\bm{\theta}}_j=\arg\max\limits_{\bm{\theta}_j\in\Theta_j}L_{n_j}(\tau_{j-1},\bm{\theta}_j)$,
and the parameter space $\Theta$ satisfies Assumption \ref{assumption3}.

Next we consider the consistency of the estimators.
It is obvious that BAR model is a typical bounded time series, which means that any finite moments of the BAR model are finite.
Based on Assumptions \ref{assumption1}-\ref{assumption3}, we can obtain the conclusion in Theorem \ref{strong} without any moment assumption.
Theorem \ref{strong} not only shows the strong consistency of the MDL procedure, but also gives the rate of convergence of the change-point estimators.
\begin{theorem}\label{strong}
{\rm\bf{(Strong Consistency)}} Let $\{X_t\}_{t=1}^n$ be observations from a piecewise stationary process (\ref{BAR}) specified by the true value vector $(m_0,\bm{\lambda}^0,\bm{\theta}^0)$ and satisfy Assumptions \ref{assumption1}-\ref{assumption3}.
The estimator $(\hat{m}_n,\hat{\bm{\lambda}}_n,\hat{\bm{\theta}}_n)$, defined in Equation (\ref{MDL_est}), is strongly consistent, i.e.,
\begin{align*}
\hat{m}_n\xrightarrow{\mathrm{a.s.}}m_0,~\hat{\bm{\lambda}}_n\xrightarrow{\mathrm{a.s.}}\bm{\lambda}^0,~\hat{\bm{\theta}}_n\xrightarrow{\mathrm{a.s.}}\bm{\theta}^0,
\end{align*}
and for each $\lambda_j^0$, $j=1,...,m_0$, there exists a $\hat{\lambda}_{j'}\in \hat{\bm{\lambda}}_n$, $1\leqslant j'\leqslant\hat{m}_n$, such that
\begin{align*}
|\hat{\lambda}_{j'}-\lambda_j^0|=o(n^{-\frac{1}{2}}),~\mathrm{a.s.},
\end{align*}
furthermore,
\begin{align}\label{theorem7.1}
\hat{\bm{\theta}}_{j'}(\hat{\lambda}_{j'-1},\hat{\lambda}_{j'})-\bm{\theta}_j^0=O_p\left(\sqrt{\frac{\log\log n}{n}}\right),
\end{align}
where
\begin{align*}
\hat{\bm{\theta}}_{j'}(\hat{\lambda}_{j'-1},\hat{\lambda}_{j'})&=\arg\max\limits_{\bm{\theta}_{j'}\in\Theta_{j'}}L_{n_{j'}}([n\hat{\lambda}_{{j'}-1}],\bm{\theta}_{j'})\\
&=\arg\max\limits_{\bm{\theta}_{j'}\in\Theta_{j'}}\left(\sum\limits_{t=[\hat{n}_{j'}\hat{\lambda}_{{j'}-1}]+1}^{[\hat{n}_{j'}\hat{\lambda}_{j'}]}\ell_t(\bm{\theta}_{j'}|X_{s},s<t)\right).
\end{align*}
\end{theorem}
\begin{remark}\label{etheta}
An $\epsilon_{\theta}>0$ in Assumption \ref{assumption3} ensures the model's sensitivity to actual change-points.
Intuitively, a larger $\epsilon_{\theta}$ emphasizes the differences between segments and may even allow approximate identification of change-points from sample path plots.
However, $\epsilon_{\theta}$ is not the sole factor affecting the efficacy of change-point estimators.
For the BAR(1) model, estimators derived from the MDL criterion exhibit varying sensitivities to changes in different parameters.
Typically, they are more sensitive to the mean parameter $p$ compared to the correlation coefficient parameter $\rho$.
Subsequent simulations in Section \ref{sect4.6} have corroborated these inferences.
\end{remark}
\begin{remark}
The $\epsilon_{\lambda}$ guarantees adequate sample between change-points, thereby validating the effectiveness of the CML estimators.
This setting is critical to optimize change-point estimators based on the MDL criterion within the framework of the CML function.
Empirical evidence from Monte Carlo studies suggests that setting $\epsilon_{\lambda}=10/n$ typically yields precise estimation outcomes.
\end{remark}

\begin{remark}\label{remark2}
Evidently, based on the findings (\ref{theorem7.1}) delineated in Theorem \ref{strong}, the convergence rate of the CML estimator $\hat{\bm{\theta}}_{n,CML}$, remains unimpacted even if the estimated segment partially extends beyond a stationary section into adjacent stationary intervals.
\end{remark}

\subsection{Searching algorithm based on genetic algorithm}\label{Set4}
In this section, we discuss the optimization algorithm of MDL criterion. Since the search space (consisting of $m$, $\bm{\tau}$ and $\bm{\theta}$) is huge, practical optimization of various IC is not a trivial task.
So far, there have been many optimization algorithms designed to solve this popular issue,
such as optimal partitioning (OP) \citep{Jackson2005}, genetic algorithm (GA) \citep{Davis2006}, pruned exact linear time (PELT) \citep{Killick2012}, pruned dynamic programming (PDP) \citep{Rigaill2010}, wild binary segmentation (WBS) \citep{Fryzlewicz2014},
just to name a few.
In this paper, GA, which is frequently used to optimize MDL criterion, is applied to solve the change-point estimation problem.

GA is a population-based search algorithm that applies the survival of the fittest concept.
Since \cite{Davis2006} proposed a classical Auto-PARM process based on GA, this type of algorithm has been widely used in optimizing MDL to identify multiple change-points.
Although the simulation and application in \cite{Davis2006} show that the Auto-PARM process is efficient, the calculation amount is quite large because the design of algorithm considers all $m$ in its parameter space $[1,M_0]$.
In fact, as we all know, MDL function is convex if we just focus on the number of change-points.
Therefore, we can obviously start the search at one end of the range $[1,M_0]$, and end the search by finding the inflection point of the MDL function. This will greatly reduce computation costs.
In view of this, we propose the following searching algorithm based on genetic algorithm (S-GA) for MCP-BAR model to identify multiple change-points.
For the purpose of clarifying the S-GA algorithm, we divide the algorithm into two parts: S step and GA step, which are given detailed introduction in the Appendix.
\begin{remark}
GA leverages the principles of natural selection and genetic mechanisms, optimizing solutions through an iterative search within the candidate solution space. Initially, the algorithm generates a set of candidate change-point positions ${\bm\tau}$, to form a population.
The fitness of each configuration is determined by the magnitude of its corresponding MDL. Through selection, crossover, and mutation processes, the algorithm continually refines the population, selecting individuals with higher fitness for reproduction, while introducing novel mutations to explore additional possible solutions. As iterations progress, the overall fitness of the population increases, converging towards the optimal or a near-optimal change-point configuration.
Due to space constraints, detailed explanations of chromosome representation, selection, crossover, mutation, and fitness function computation are omitted and can be found in \cite{Davis2006} and \cite{Sheng2023}.
\end{remark}

\section{Simulation}
In this section, our target is to investigate the performances of the CUSUM test and the S-GA algorithm for the detection and estimation of the change-points.
\subsection{CUSUM test}
Some simulations are conducted to investigate the performances of the CUSUM test.
Also, the test statistic $C_{n}^{\text{Zhang}}$, proposed in Section 3.3 of \cite{Zhang2023}, is considered as a comparison.
We select the significance level $\gamma=0.01,0.05$, the associated critical value is 3.269 and 2.408 \citep{Lee2003}, sample size $n=200,500,1000$, $N=10$ and $k_{0}=10$.
All results are summarized in Tables \ref{test_table1}-\ref{test_table3} based on 1000 replications.
For analyzing the empirical size, the data is generated from the BAR(1) model with three parameter combinations:\\
Model T$_{1}$: $(\rho,p)=(-0.1,0.6)$;\\
Model T$_{2}$: $(\rho,p)=(0.1,0.3)$;\\
Model T$_{3}$: $(\rho,p)=(0.4,0.3)$.\\
Then, in term of the empirical power, we consider the following three classes of models, which are corresponding to Models T$_{1}$-T$_{3}$:
\vspace{1.5mm}\\
Model T$_{11}$, only $\rho$ change: $(\rho,p)=(-0.1,0.6)$ change to $(\rho,p)=(0.5,0.6)$ at $\tau=0.5n$;\\
Model T$_{12}$, only $p$ change: $(\rho,p)=(-0.1,0.6)$ change to $(\rho,p)=(-0.1,0.3)$ at $\tau=0.5n$;\\
Model T$_{13}$: $(\rho,p)=(-0.1,0.6)$ change to $(\rho,p)=(0.1,0.3)$ at $\tau=0.5n$.
\vspace{2mm}\\
Model T$_{21}$, only $\rho$ change: $(\rho,p)=(0.1,0.3)$ change to $(\rho,p)=(0.5,0.3)$ at $\tau=0.5n$;\\
Model T$_{22}$, only $p$ change: $(\rho,p)=(0.1,0.3)$ change to $(\rho,p)=(0.1,0.6)$ at $\tau=0.5n$;\\
Model T$_{23}$: $(\rho,p)=(0.1,0.3)$ change to $(\rho,p)=(0.3,0.5)$ at $\tau=0.5n$.
\vspace{2mm}\\
Model T$_{31}$, only $\rho$ change: $(\rho,p)=(0.4,0.3)$ change to $(\rho,p)=(-0.2,0.3)$ at $\tau=0.5n$;\\
Model T$_{32}$, only $p$ change: $(\rho,p)=(0.4,0.3)$ change to $(\rho,p)=(0.4,0.6)$ at $\tau=0.5n$;\\
Model T$_{33}$: $(\rho,p)=(0.4,0.3)$ change to $(\rho,p)=(-0.2,0.6)$ at $\tau=0.5n$.

The results are reported in Tables \ref{test_table1}-\ref{test_table3}, which show that the empirical sizes are close to the significant levels $\gamma=0.05,0.1$, as expected.
Although, the proposed test statistics give satisfactory performances for the empirical sizes (especially the CUSUM test based on the MQL and CML estimators), the $C_n^{\text{Zhang}}$ statistic achieves convergence to the significance level more rapidly, i.e., $C_n^{\text{Zhang}}$ performs better in terms of the empirical sizes.
Regarding empirical power, as evidenced by Tables \ref{test_table1}-\ref{test_table3}, the $C_n^{CLS}$ and $C_n^{\text{Zhang}}$ statistics exhibit limited sensitivity in detecting changes in the autocorrelation structure, especially under conditions of modest sample sizes and isolated variations in $\rho$ (refer to the results for Models T$_{11}$, T$_{21}$, T$_{31}$ across Tables \ref{test_table1}-\ref{test_table3}).
However, the $C_n^{CLS}$ statistic demonstrates a notably enhanced empirical power compared to $C_n^{\text{Zhang}}$.
In scenarios with a constant $p$ value, such as Models T$_{11}$, T$_{21}$, T$_{31}$, the $C_n^{\text{Zhang}}$ statistic remains small, even as the sample size increases.
Conversely, the proposed statistics progressively approximate to 1 with increasing sample size, with $C_n^{CML}$ performing best, followed by $C_n^{MQL}$.
This trend is anticipated, given that the CML estimators more accurately account for the probabilistic distribution characteristics of the data, thereby facilitating more precise change-point detection.
Consequently, it is inferred that the proposed test statistics are predominantly effective in identifying mean shifts over alterations in the autocorrelation coefficient.
For practical applications, the CML-based and MQL-based test statistic are recommended.
\begin{table}
\centering
\caption{The empirical size and power for CUSUM test for Models T$_1$-T$_{13}$.}\label{test_table1}
\vspace{1mm}
\begin{tabular}{*{8}{c}}
\toprule
\multirow{2}*{Size}&&\multicolumn{3}{c}{$\gamma=0.01$}&\multicolumn{3}{c}{$\gamma=0.05$}\\\cmidrule(lr){3-5}\cmidrule(lr){6-8}
&Sample size&200&500&1000&200&500&1000\\\cmidrule(lr){1-1}\cmidrule(lr){2-8}
T$_1$&$C_n^{CLS}$&0.0450 &0.0280 &0.0180 &0.1190 &0.0830 &0.0870 \\
{ with $(\rho,p)=(-0.1,0.6)$}&$C_n^{MQL}$&0.0110 &0.0110 &0.0130 &0.0410 &0.0620 &0.0630 \\
&$C_n^{CML}$&0.0350 &0.0130 &0.0090 &0.0590 &0.0540 &0.0450 \\
&$C_n^{\text{Zhang}}$&0.0100 &0.0080 &0.0090 &0.0330 &0.0460 &0.0440 \\\midrule
\multirow{2}*{Power}&&\multicolumn{3}{c}{$\gamma=0.01$}&\multicolumn{3}{c}{$\gamma=0.05$}\\\cmidrule(lr){3-5} \cmidrule(lr){6-8}
&Sample size&200&500&1000&200&500&1000\\\cmidrule(lr){1-1}\cmidrule(lr){2-8}
T$_{11}$&$C_n^{CLS}$&0.3610 &0.9890 &1.0000 &0.6220 &1.0000 &1.0000 \\
only $\rho$&$C_n^{MQL}$&0.8150 &0.9990 &1.0000 &0.9200 &1.0000 &1.0000 \\
change to $0.5$&$C_n^{CML}$&0.8300 &0.9990 &1.0000 &0.9290 &1.0000 &1.0000 \\
&$C_n^{\text{Zhang}}$&0.0330 &0.0510 &0.0480 &0.1000 &0.1090 &0.1280 \\\cmidrule(lr){3-8}
T$_{12}$&$C_n^{CLS}$&1.0000 &1.0000 &1.0000 &1.0000 &1.0000 &1.0000 \\
only $p$&$C_n^{MQL}$&1.0000 &1.0000 &1.0000 &1.0000 &1.0000 &1.0000 \\
change to $0.3$&$C_n^{CML}$&1.0000 &1.0000 &1.0000 &1.0000 &1.0000 &1.0000 \\
&$C_n^{\text{Zhang}}$&1.0000 &1.0000 &1.0000 &1.0000 &1.0000 &1.0000 \\\cmidrule(lr){3-8}
T$_{13}$&$C_n^{CLS}$&1.0000 &1.0000 &1.0000 &1.0000 &1.0000 &1.0000 \\
$(\rho,p)$&$C_n^{MQL}$&1.0000 &1.0000 &1.0000 &1.0000 &1.0000 &1.0000 \\
change to $(0.1,0.3)$&$C_n^{CML}$&1.0000 &1.0000 &1.0000 &1.0000 &1.0000 &1.0000 \\
&$C_n^{\text{Zhang}}$&1.0000 &1.0000 &1.0000 &1.0000 &1.0000 &1.0000 \\
\bottomrule
\end{tabular}
\end{table}

\begin{table}
\centering
\caption{The empirical size and power for CUSUM test for Models T$_2$-T$_{23}$.}\label{test_table2}
\vspace{1mm}
\begin{tabular}{*{8}{c}}
\toprule
\multirow{2}*{Size}&&\multicolumn{3}{c}{$\gamma=0.01$}&\multicolumn{3}{c}{$\gamma=0.05$}\\\cmidrule(lr){3-5}\cmidrule(lr){6-8}
&Sample size&200&500&2000&200&500&1000\\\cmidrule(lr){1-1}\cmidrule(lr){2-8}
T$_2$&$C_n^{CLS}$&0.0180 &0.0070 &0.0200 &0.0480 &0.0420 &0.0540 \\
{ with $(\rho,p)=(0.1,0.3)$}&$C_n^{MQL}$&0.0220 &0.0140 &0.0220 &0.0640 &0.0600 &0.0600 \\
&$C_n^{CML}$&0.0340 &0.0060 &0.0120 &0.0780 &0.0370 &0.0420 \\
&$C_n^{\text{Zhang}}$&0.0080 &0.0110 &0.0120 &0.0380 &0.0360 &0.0470 \\\midrule
\multirow{2}*{Power}&&\multicolumn{3}{c}{$\gamma=0.01$}&\multicolumn{3}{c}{$\gamma=0.05$}\\\cmidrule(lr){3-5} \cmidrule(lr){6-8}
&Sample size&200&500&1000&200&500&1000\\\cmidrule(lr){1-1}\cmidrule(lr){2-8}
T$_{21}$&$C_n^{CLS}$&0.3330 &0.8600 &0.9970 &0.5310 &0.9460 &0.9990 \\
only $\rho$&$C_n^{MQL}$&0.4730 &0.9050 &0.9970 &0.6640 &0.9640 &0.9990 \\
change to $0.5$&$C_n^{CML}$&0.6980 &0.9740 &1.0000 &0.8430 &0.9940 &1.0000 \\
&$C_n^{\text{Zhang}}$&0.0130 &0.0190 &0.0210 &0.0660 &0.0730 &0.0880 \\\cmidrule(lr){3-8}
T$_{22}$&$C_n^{CLS}$&1.0000 &1.0000 &1.0000 &1.0000 &1.0000 &1.0000 \\
only $p$&$C_n^{MQL}$&1.0000 &1.0000 &1.0000 &1.0000 &1.0000 &1.0000 \\
change to $0.6$&$C_n^{CML}$&1.0000 &1.0000 &1.0000 &1.0000 &1.0000 &1.0000 \\
&$C_n^{\text{Zhang}}$&1.0000 &1.0000 &1.0000 &1.0000 &1.0000 &1.0000 \\\cmidrule(lr){3-8}
T$_{23}$&$C_n^{CLS}$&1.0000 &1.0000 &1.0000 &1.0000 &1.0000 &1.0000 \\
$(\rho,p)$&$C_n^{MQL}$&1.0000 &1.0000 &1.0000 &1.0000 &1.0000 &1.0000 \\
change to $(0.3,0.5)$&$C_n^{CML}$&1.0000 &1.0000 &1.0000 &1.0000 &1.0000 &1.0000 \\
&$C_n^{\text{Zhang}}$&1.0000 &1.0000 &1.0000 &1.0000 &1.0000 &1.0000 \\
\bottomrule
\end{tabular}
\end{table}

\begin{table}
\centering
\caption{The empirical size and power for CUSUM test for Models T$_3$-T$_{33}$.}\label{test_table3}
\vspace{1mm}
\begin{tabular}{*{8}{c}}
\toprule
\multirow{2}*{Size}&&\multicolumn{3}{c}{$\gamma=0.01$}&\multicolumn{3}{c}{$\gamma=0.05$}\\\cmidrule(lr){3-5}\cmidrule(lr){6-8}
&Sample size&200&500&1000&200&500&1000\\\cmidrule(lr){1-1}\cmidrule(lr){2-8}
T$_3$&$C_n^{CLS}$&0.0290 &0.0200 &0.0180 &0.0730 &0.0590 &0.0650 \\
{ with $(\rho,p)=(0.4,0.3)$}&$C_n^{MQL}$&0.0390 &0.0220 &0.0220 &0.1090 &0.0710 &0.0560 \\
&$C_n^{CML}$&0.0370 &0.0040 &0.0050 &0.0170 &0.0200 &0.0340 \\
&$C_n^{\text{Zhang}}$&0.0070 &0.0100 &0.0090 &0.0360 &0.0340 &0.0420 \\\midrule
\multirow{2}*{Power}&&\multicolumn{3}{c}{$\gamma=0.01$}&\multicolumn{3}{c}{$\gamma=0.05$}\\\cmidrule(lr){3-5} \cmidrule(lr){6-8}
&Sample size&200&500&1000&200&500&1000\\\cmidrule(lr){1-1}\cmidrule(lr){2-8}
T$_{31}$&$C_n^{CLS}$&0.9890 &1.0000 &1.0000 &0.9980 &1.0000 &1.0000 \\
only $\rho$&$C_n^{MQL}$&0.9890 &1.0000 &1.0000 &0.9980 &1.0000 &1.0000 \\
change to $-0.2$&$C_n^{CML}$&0.9990 &1.0000 &1.0000 &1.0000 &1.0000 &1.0000 \\
&$C_n^{\text{Zhang}}$&0.8680 &1.0000 &1.0000 &0.9520 &1.0000 &1.0000 \\\cmidrule(lr){3-8}
T$_{32}$&$C_n^{CLS}$&1.0000 &1.0000 &1.0000 &1.0000 &1.0000 &1.0000 \\
only $p$&$C_n^{MQL}$&1.0000 &1.0000 &1.0000 &1.0000 &1.0000 &1.0000 \\
change to $0.6$&$C_n^{CML}$&1.0000 &1.0000 &1.0000 &1.0000 &1.0000 &1.0000 \\
&$C_n^{\text{Zhang}}$&1.0000 &1.0000 &1.0000 &1.0000 &1.0000 &1.0000 \\\cmidrule(lr){3-8}
T$_{33}$&$C_n^{CLS}$&1.0000 &1.0000 &1.0000 &1.0000 &1.0000 &1.0000 \\
$(\rho,p)$&$C_n^{MQL}$&1.0000 &1.0000 &1.0000 &1.0000 &1.0000 &1.0000 \\
change to $(-0.2,0.6)$&$C_n^{CML}$&1.0000 &1.0000 &1.0000 &1.0000 &1.0000 &1.0000 \\
&$C_n^{\text{Zhang}}$&1.0000 &1.0000 &1.0000 &1.0000 &1.0000 &1.0000 \\
\bottomrule
\end{tabular}
\end{table}

\subsection{Change-point estimation}
To evaluate the finite-sample performance of the proposed S-GA algorithm, we conduct extensive simulation studies, which are split into four parts.
First, we consider the sensitivity of genetic algorithm to tunning parameter CF.
Second, we compare the performance in minimizing MDL based on the CML function evaluated on the CLS and CML estimators.
Third, S-GA algorithm was compared with GA algorithm \citep{Davis2006}.
In the last part, simulations are utilized to study the consistency conclusion in Theorem \ref{strong}.

In this section, we not only calculate the correct rate of the number of change-points, CR($m$)$=1/r\sum_{i=1}^r{\rm I}(\hat{m}_n^{(i)}=m_0)$,
where ${\rm I}(A)$ denotes the indicator function, assigning a value of $1$ if $A$ is true, and zero otherwise, $r$ is the number of repetitions, $\hat{m}_n^{(i)}$ is the $i$-th estimator for $m_0$,
but also report the following two type evaluation metrics to measure the performance of the change-points location estimator $\hat{\bm{\lambda}}_n$:
\begin{align*}
\zeta(\bm{\lambda}^0|\hat{\bm{\lambda}}_n)=\sup\limits_{b\in\hat{\bm{\lambda}}_n}\inf\limits_{a\in\bm{\lambda}^0}|a-b|,
~~~\zeta(\hat{\bm{\lambda}}_n|\bm{\lambda}^0)=\sup\limits_{b\in\bm{\lambda}^0}\inf\limits_{a\in\hat{\bm{\lambda}}_n}|a-b|,~\text{\citep{Boysen2009}},
\end{align*}
which quantify the under-segmentation error and the over-segmentation error, respectively. A desirable estimator should be able to balance both quantities. In addition, to evaluate the location accuracy of the estimated change-points, we also report the following distance from the estimated set $\hat{\bm{\lambda}}_n$ and true change-points set $\bm{\lambda}^0$: 
$$d(\hat{\bm{\lambda}}_n,\bm{\lambda}^0)=\frac{1}{|\bm{\lambda}^0|}\sum\limits_{\lambda_k^0\in\bm{\lambda}^0}\min\limits_{\hat{\lambda}_j\in\hat{\bm{\lambda}}_n}|\hat{\lambda}_j-\lambda_k^0|,~\mbox{\citep{Chen2021}}.$$
Also, the empirical biases (Bias) and mean square errors (MSE) for every estimator is considered  when $m$ is correctly estimated. All simulations are carried out using the MATLAB software. The empirical results displayed in the tables are computed over $1000$ replications.

We consider two classes of scenarios with two change-points (A-type) and three change-points (B-type) in our simulation study.
For two change-points case or three change-points case,
we also investigate three types of changes in mean and correlation:
(1) the mean is a constant but the autocorrelation coefficient changes (A1, B1); (2) the autocorrelation coefficient is a constant but the mean changes (A2, B2); (3) both mean and autocorrelation coefficient change (A3, B3).
The parameter settings, sample size, and the location of change-points are specified in Tables \ref{model_set_2} and \ref{model_set_3}.
\begin{table}[H]
\centering
\caption{Parameter combinations for Models (A1)-(A3) (upper bound $N=10$).}\label{model_set_2}
\begin{tabular}{*{2}{c}rrr*{3}{c}}
\toprule
\multirow{2}*{Model}&&\multicolumn{3}{c}{Two change}&&\multicolumn{2}{c}{Change-point location}\\ \cmidrule(l){3-5}\cmidrule(l){7-8}
&Segment&I&II&III&Sample size&$\tau_1$&$\tau_2$\\\cmidrule(lr){1-1}\cmidrule(l){2-5}\cmidrule(l){6-8}
A1&$p$&0.5&0.5&0.5&200&70&140\\
$\rho$ change, $p$ same&$\rho$&$-$0.2&0.6&0.1&500&150&350\\\cmidrule(l){3-5}
A2&$p$&0.3&0.5&0.7&800&300&450\\
$\rho$ same, $p$ change&$\rho$&0.2&0.2&0.2&&&\\\cmidrule(l){3-5}
A3&$p$&0.3&0.5&0.7&&&\\
$(\rho,p)$ change&$\rho$&$-$0.2&0.6&0.3&&&\\
\bottomrule
\end{tabular}
\end{table}

\begin{table}[H]
\centering
\caption{Parameter combinations for Models (A1)-(A3) (upper bound $N=10$).}\label{model_set_3}
\begin{tabular}{*{2}{c}rrrr*{4}{c}}
\toprule
\multirow{2}*{Model}&&\multicolumn{4}{c}{Three change}&&\multicolumn{3}{c}{Change-points location}\\ \cmidrule(l){3-6}\cmidrule(l){8-10}
&Segment&I&II&III&IV&Sample size&$\tau_1$&$\tau_2$&$\tau_3$\\\cmidrule(lr){1-1}\cmidrule(l){2-6}\cmidrule(l){7-10}
B1&$p$&0.5&0.5&0.5&0.5&200&50&100&150\\
$\rho$ change, $p$ same&$\rho$&$-$0.2&0.6&0.1&0.4&500&100&225&390\\\cmidrule(l){3-6}
B2&$p$&0.2&0.4&0.6&0.8&800&200&400&650\\
$\rho$ same, $p$ change&$\rho$&0.3&0.3&0.3&0.3&&&&\\\cmidrule(l){3-6}
B3&$p$&0.3&0.4&0.6&0.8&&&&\\
$(\rho,p)$ change &$\rho$&$-$0.2&$-$0.1&0.2&0.4&&&&\\
\bottomrule
\end{tabular}
\end{table}

\subsubsection{Sensitivity analysis for the tunning parameter CF}
In the S-GA algorithm, it is necessary to measure the influence of the setting of tunning parameter CF (the option in "ga" function) on estimation effect.
For this, we set CF$=0.3,0.55,0.8$ and sample size $n=200$ and compare the estimation effect for Models (A1)-(A3).
The simulation results are summarized in Table \ref{sensitive_cf}.
From the CR($m$) results, CF$=0.3$ almost attains the value closest to 1, followed by CF$=0.55$, though the difference between them is very small.
Regarding the under-segmentation error and the over-segmentation error, CF$=0.55$ achieves a balance between these two types of errors, with $\zeta(\bm{\lambda}^0|\hat{\bm{\lambda}}_n)$ and $\zeta(\hat{\bm{\lambda}}_n|\bm{\lambda}^0)$ being not significantly different, and its performance is noticeably better than that of CF$=0.3, 0.8$.
In terms of estimation accuracy index $d(\hat{\bm{\lambda}}_n,\bm{\lambda}^0)$, CF$=0.8$ shows the best performance, though CF$=0.55$ is not far behind.
Hence, we set CF$=0.55$ in the simulations.

\begin{table}
	\centering
	\caption{Sensitivity analysis for CF with sample size $n=200$.}\label{sensitive_cf}
	\begin{tabular}{*{6}{c}rr}
\toprule
\multicolumn{8}{c}{CF$=0.3$}\\\cmidrule(lr){1-8}
Model&CR($m$)&$\zeta(\bm{\lambda}^0|\hat{\bm{\lambda}}_n)$&$\zeta(\hat{\bm{\lambda}}_n|\bm{\lambda}^0)$&$d(\hat{\bm{\lambda}}_n,\bm{\lambda}^0)$&&\multicolumn{1}{c}{$\lambda_1$}&\multicolumn{1}{c}{$\lambda_2$}\\\cmidrule(lr){1-1}\cmidrule(lr){2-2}\cmidrule(lr){3-5}\cmidrule(lr){7-8}
A1&0.545&0.0687 &0.2126 &0.3220 &Bias&$-$0.0114&0.0102 \\
&&&&&MSE&0.0029 &0.0086 \\
A2&0.924&0.0404 &0.0386 &0.0606 &Bias&0.0015 &$-$0.0034\\
&&&&&MSE&0.0014 &0.0013 \\
A3&0.874&0.0407 &0.0620 &0.0914 &Bias&$-$0.0046&0.0011 \\
&&&&&MSE&0.0008 &0.0025 \\\hline
\multicolumn{8}{c}{CF$=0.55$}\\\cmidrule(lr){1-8}
Model&CR($m$)&$\zeta(\bm{\lambda}^0|\hat{\bm{\lambda}}_n)$&$\zeta(\hat{\bm{\lambda}}_n|\bm{\lambda}^0)$&$d(\hat{\bm{\lambda}}_n,\bm{\lambda}^0)$&&\multicolumn{1}{c}{$\lambda_1$}&\multicolumn{1}{c}{$\lambda_2$}\\\cmidrule(lr){1-1}\cmidrule(lr){2-2}\cmidrule(lr){3-5}\cmidrule(lr){7-8}
A1&0.547&0.0748 &0.1398 &0.2130 &Bias&$-$0.0056&0.0018 \\
&&&&&MSE&0.0027 &0.0081 \\
A2&0.917&0.0434 &0.0326 &0.0527 &Bias&$-$0.0002&$-$0.0061\\
&&&&&MSE&0.0010 &0.0010 \\
A3&0.871&0.0452 &0.0400 &0.0628 &Bias&$-$0.0042&0.0018 \\
&&&&&MSE&0.0008 &0.0028 \\\hline
\multicolumn{8}{c}{CF$=0.8$}\\\cmidrule(lr){1-8}
Model&CR($m$)&$\zeta(\bm{\lambda}^0|\hat{\bm{\lambda}}_n)$&$\zeta(\hat{\bm{\lambda}}_n|\bm{\lambda}^0)$&$d(\hat{\bm{\lambda}}_n,\bm{\lambda}^0)$&&\multicolumn{1}{c}{$\lambda_1$}&\multicolumn{1}{c}{$\lambda_2$}\\\cmidrule(lr){1-1}\cmidrule(lr){2-2}\cmidrule(lr){3-5}\cmidrule(lr){7-8}
A1&0.523&0.0785 &0.1083 &0.1694 &Bias&$-$0.0060&0.0062 \\
&&&&&MSE&0.0022 &0.0070 \\
A2&0.924&0.0438 &0.0326 &0.0527 &Bias&$-$0.0001&$-$0.0058\\
&&&&&MSE&0.0011 &0.0010 \\
A3&0.853&0.0501 &0.0382 &0.0600 &Bias&$-$0.0022&0.0024 \\
&&&&&MSE&0.0007 &0.0028 \\
\bottomrule
	\end{tabular}
\end{table}

\subsubsection{Comparing S-GA based on CML estimator with CLS estimator}
In this section, our main goal is to determine whether the settings outlined in Setting tips (1) are reasonable.
We utilized the S-GA algorithm with both the CML and CLS estimators to estimate the number and locations of change-points.
We summarized the assessment criterion and durations in Table \ref{clscmltime}, which shows that the following four criterion: the two estimators in terms of CR($m$), the under-segmentation error $\zeta(\bm{\lambda}^0|\hat{\bm{\lambda}}_n)$, the over-segmentation error $\zeta(\hat{\bm{\lambda}}_n|\bm{\lambda}^0)$, and the accuracy of the estimated change-point locations $d(\hat{\bm{\lambda}}_n,\bm{\lambda}^0)$, give similar results based on the CLS and CML estimators.
Although the CML estimator performs slightly better than the CLS estimator in several criterion,
the durations based on the CML method are much longer than those based on the CLS method.
Therefore, based on overall performance, it is a reasonable choice to use the CLS estimator rather than the CML estimator.

\begin{table}[H]
	\scriptsize
\centering
	\caption{Comparision of S-GA algorithm based on CML and CLS estimators}\label{clscmltime}
	\begin{tabular}{cccccccrrc}
\toprule
\multicolumn{10}{c}{CLS}\\\cmidrule(lr){1-10}
Model&Sample size&CR($m$)&$\zeta(\bm{\lambda}^0|\hat{\bm{\lambda}}_n)$&$\zeta(\hat{\bm{\lambda}}_n|\bm{\lambda}^0)$&$d(\hat{\bm{\lambda}}_n,\bm{\lambda}^0)$&&\multicolumn{1}{c}{$\lambda_1$}&\multicolumn{1}{c}{$\lambda_2$}&Duration(s)\\\cmidrule(lr){1-1}\cmidrule(lr){2-2}\cmidrule(lr){3-3}\cmidrule(lr){4-6}\cmidrule(lr){8-9}\cmidrule(lr){10-10}
A1&$n=$200&0.540 &0.0707 &0.2113 &0.3225 &Bias&$-$0.0083&0.0153 &2941.085 \\
&&&&&&MSE&0.0026 &0.0071 &\\
&$n=$500&0.936 &0.0316 &0.0385 &0.0599 &Bias&$-$0.0026&0.0099 &5245.813 \\
&&&&&&MSE&0.0003 &0.0013 &\\
&$n=$800&0.947 &0.0247 &0.0259 &0.0462 &Bias&$-$0.0016&0.0087 &7374.188 \\
&&&&&&MSE&0.0002 &0.0012 &\\\cmidrule(lr){1-10}
A2&$n=$200&0.910 &0.0390 &0.0392 &0.0615 &Bias&$-$0.0004&$-$0.0045&3218.818 \\
&&&&&&MSE&0.0010 &0.0011 &\\
&$n=$500&0.970 &0.0146 &0.0109 &0.0183 &Bias&$-$0.0008&$-$0.0017&5008.807 \\
&&&&&&MSE&0.0001 &0.0001 &\\
&$n=$800&0.984 &0.0090 &0.0066 &0.0126 &Bias&$-$0.0005&$-$0.0013&6180.803 \\
&&&&&&MSE&0.0000 &0.0000 &\\\midrule
\multicolumn{10}{c}{CML}\\\cmidrule(lr){1-10}
Model&Sample size&CR($m$)&$\zeta(\bm{\lambda}^0|\hat{\bm{\lambda}}_n)$&$\zeta(\hat{\bm{\lambda}}_n|\bm{\lambda}^0)$&$d(\hat{\bm{\lambda}}_n,\bm{\lambda}^0)$&&\multicolumn{1}{c}{$\lambda_1$}&\multicolumn{1}{c}{$\lambda_2$}&Duration(s)\\\cmidrule(lr){1-1}\cmidrule(lr){2-2}\cmidrule(lr){3-3}\cmidrule(lr){4-6}\cmidrule(lr){8-9}\cmidrule(lr){10-10}
A1&$n=$200&0.533 &0.0782 &0.2084 &0.3204 &Bias&$-$0.0085&0.0168 &154358.245 \\
&&&&&&MSE&0.0030 &0.0079 &\\
&$n=$500&0.929 &0.0333 &0.0384 &0.0601 &Bias&$-$0.0021&0.0103 &382916.025 \\
&&&&&&MSE&0.0003 &0.0014 &\\
&$n=$800&0.954 &0.0254 &0.0219 &0.0398 &Bias&$-$0.0016&0.0082 &464652.567 \\
&&&&&&MSE&0.0002 &0.0010 &\\\cmidrule(lr){1-10}
A2&$n=$200&0.873 &0.0434 &0.0381 &0.0598 &Bias&$-$0.0003&$-$0.0045&185618.739 \\
&&&&&&MSE&0.0010 &0.0012 &\\
&$n=$500&0.956 &0.0160 &0.0109 &0.0184 &Bias&$-$0.0006&$-$0.0018&309415.621 \\
&&&&&&MSE&0.0001 &0.0001 &\\
&$n=$800&0.980 &0.0099 &0.0069 &0.0130 &Bias&$-$0.0006&$-$0.0014&400169.839 \\
&&&&&&MSE&0.0000 &0.0001 &\\
\bottomrule
\end{tabular}
\end{table}

\subsubsection{Comparison between the Auto-PARM and S-GA}
In this section, we conduct some simulations to illustrate that the proposed S-GA algorithm significantly improves the efficiency
without the loss of estimation accuracy.
For this, we compare the S-GA algorithm with the Auto-PARM proposed by \cite{Davis2006} and the simulation results are given in Table \ref{cls_cml_time}.
We can see that the S-GA algorithm outperforms the Auto-PARM when we consider estimation accuracy and computational cost.
From the duration in Table \ref{cls_cml_time}, we can see that the S-GA algorithm is 60 times faster than the Auto-PARM when we set the sample size $n=200$ and this advantage becomes more apparent as the sample size increases.
\begin{table}[H]
\scriptsize
\centering
\caption{Comparison between the Auto-PARM and S-GA}\label{cls_cml_time}
\begin{tabular}{*{7}{c}rrc}
\toprule
\multicolumn{10}{c}{Auto-PARM-Davis}\\\cmidrule(lr){1-10}
Model&Sample size&CR(m)&$\zeta(\bm{\lambda}^0|\hat{\bm{\lambda}}_n)$&$\zeta(\hat{\bm{\lambda}}_n|\bm{\lambda}^0)$&$d(\hat{\bm{\lambda}}_n,\bm{\lambda}^0)$&&\multicolumn{1}{c}{$\lambda_1$}&\multicolumn{1}{c}{$\lambda_2$}&Duration(s)\\\cmidrule(lr){1-1}\cmidrule(lr){2-2}\cmidrule(lr){3-3}\cmidrule(lr){4-6}\cmidrule(lr){8-9}\cmidrule(lr){10-10}
A1&$n=$200&0.535 &0.0695 &0.2090 &0.3181 &Bias&$-$0.0037&0.0144 &162325.172\\
&&&&&&MSE&0.0025 &0.0058 &\\
&$n=$500&0.929 &0.0335 &0.0427 &0.0659 &Bias&$-$0.0016&0.0104 &533674.816\\
&&&&&&MSE&0.0004 &0.0016 &\\
&$n=$800&0.950 &0.0315 &0.0230 &0.0414 &Bias&$-$0.0011&0.0062 &1187773.597\\
&&&&&&MSE&0.0002 &0.0007 &\\\hline
\multicolumn{10}{c}{S-GA}\\\cmidrule(lr){1-10}
Model&Sample size&CR(m)&$\zeta(\bm{\lambda}^0|\hat{\bm{\lambda}}_n)$&$\zeta(\hat{\bm{\lambda}}_n|\bm{\lambda}^0)$&$d(\hat{\bm{\lambda}}_n,\bm{\lambda}^0)$&&\multicolumn{1}{c}{$\lambda_1$}&\multicolumn{1}{c}{$\lambda_2$}&Duration(s)\\\cmidrule(lr){1-1}\cmidrule(lr){2-2}\cmidrule(lr){3-3}\cmidrule(lr){4-6}\cmidrule(lr){8-9}\cmidrule(lr){10-10}
A1&$n=$200&0.540 &0.0721 &0.2056 &0.3141 &Bias&$-$0.0083&0.0153 &2941.085\\
&&&&&&MSE&0.0026 &0.0071 &\\
&$n=$500&0.936 &0.0330 &0.0376 &0.0585 &Bias&$-$0.0026&0.0099 &5245.813\\
&&&&&&MSE&0.0003 &0.0013 &\\
&$n=$800&0.947 &0.0336 &0.0227 &0.0408 &Bias&$-$0.0016&0.0087 &7374.188\\
&&&&&&MSE&0.0002 &0.0012 &\\
\bottomrule
\end{tabular}
\end{table}

\subsubsection{Consistency Analysis}\label{sect4.6}
The previous section primarily focused on the effectiveness and competitiveness of the algorithm. In this section, we mainly consider the consistency of the number and location of change-points, and the parameters in each segment under the MDL criterion (see Theorem \ref{strong}).
The corresponding results are summarized in Tables \ref{change_point_number_rate_2}-\ref{parameter estimate_Estimate_tau}.

Firstly, from Tables \ref{change_point_number_rate_2}-\ref{change_point_number_rate_3}, it can be seen that the accuracy rate of estimated number of change-points are overall satisfactory.
Although in the case of constant mean and small sample size (Models (A1) and (B1) with $n=200$), the accuracy rate of estimated number of change-points has large deviation, but it increasingly approaches 1 as the sample size increases.
Furthermore, comparing the results of Models (A1) and (A2) in Table \ref{change_point_number_rate_2}, despite the minimum parameter distance in Model (A1) being $\epsilon_{\theta}=0.25$ and in Model (A2) being $\epsilon_{\theta}=0.04$, the change-point estimation performance under Model (A2) is noticeably superior to that under Model (A1).
This validates the conclusion stated in Remark \ref{etheta}, that is, the change-point estimators are more sensitive to the mean parameter $p$ compared to the correlation coefficient parameter $\rho$.

Secondly, from the results of the change-point location estimation in Table \ref{2change_point}, as the sample size increases, the accuracy $d(\hat{\bm{\lambda}}_n,\bm{\lambda}^0)$ gives better performance, and the estimation results of the under-segmentation error and the over-segmentation error tend to balance.

Thirdly, Table \ref{parameter estimate_Estimate_tau} is designed to verify Remark \ref{remark2}, that is, the convergence of the estimator $\hat{\bm{\theta}}$ is not affected even when the estimated piece may not be fully inside a stationary piece of a time series but involves part of the adjacent stationary pieces. We summarize the results of Models (A1)-(A3) based on the true change-point location and the estimated change-point location. Although the parameter estimation results based on the true change-point location are better than those based on the estimated change-point location, as the sample size increases, both estimators are consistent. This fully conforms to the conclusion we obtained, that even if the estimated segment is not stationary, it does not affect the convergence of the estimators.

\begin{table}[H]
	\centering
	\caption{Summary of the number of estimated change-points (two change-points case)
}\label{change_point_number_rate_2}
	\begin{tabular}{*{6}{c}}
\toprule
&&\multicolumn{4}{c}{Number of segments($m$)}\\
&&\multicolumn{4}{c}{Change-points($\%$)}\\\cmidrule(l){3-6}
Model&Sample size&$m=1$&$m=2$&$m=3$&$m>3$\\\cmidrule(lr){1-1}\cmidrule(lr){2-2}\cmidrule(lr){3-6}
A1&$n=$200&0.427&\textbf{0.54}&0.031&0.002\\
&$n=$500&0.027&\textbf{0.936}&0.037&0\\
&$n=$800&0.029&\textbf{0.947}&0.024&0\\\cmidrule(lr){1-6}
A2&$n=$200&0.025&\textbf{0.91}&0.064&0.001\\
&$n=$500&0&\textbf{0.97}&0.03&0\\
&$n=$800&0&\textbf{0.984}&0.016&0\\\cmidrule(lr){1-6}
A3&$n=$200&0.086&\textbf{0.857}&0.057&0\\
&$n=$500&0&\textbf{0.976}&0.024&0\\
&$n=$800&0&\textbf{0.98}&0.02&0\\
\bottomrule
	\end{tabular}
\end{table}

\begin{table}[H]
	\centering
	\caption{Summary of the number of estimated change-points (three change-points case)}\label{change_point_number_rate_3}
	\begin{tabular}{*{7}{c}}
\toprule
&&\multicolumn{5}{c}{Number of segments($m$)}\\
&&\multicolumn{5}{c}{Change-points($\%$)}\\\cmidrule(l){3-7}
Model&Sample size&$m=1$&$m=2$&$m=3$&$m=4$&$m>4$\\\cmidrule(lr){1-1}\cmidrule(lr){2-2}\cmidrule(lr){3-7}
B1&$n=$200&0.697&0.129&\textbf{0.168}&0.006&0\\
&$n=$500&0.257&0.046&\textbf{0.671}&0.025&0.001\\
&$n=$800&0.06&0.036&\textbf{0.886}&0.018&0\\\cmidrule(l){1-7}
B2&$n=$200&0.008&0.316&\textbf{0.647}&0.029&0\\
&$n=$500&0&0.02&\textbf{0.947}&0.032&0.001\\
&$n=$800&0&0.007&\textbf{0.979}&0.014&0\\\cmidrule(l){1-7}
B3&$n=$200&0.022&0.371&\textbf{0.588}&0.019&0\\
&$n=$500&0&0.105&\textbf{0.861}&0.033&0.001\\
&$n=$800&0&0.038&\textbf{0.942}&0.02&0\\
\bottomrule
\end{tabular}
\end{table}

\begin{table}[H]
	\scriptsize
	\caption{Summary of estimated location of change-points for Models (A1)-(A3).}\label{2change_point}
\begin{adjustwidth}{2cm}{-1cm}
	\begin{tabular}{*{3}{c}rrr*{3}{c}}
\toprule
Model&Sample size&&\multicolumn{1}{c}{$\lambda_1$}&\multicolumn{1}{c}{$\lambda_2$}&&$\zeta(\bm{\lambda}^0|\hat{\bm{\lambda}}_n)$&$\zeta(\hat{\bm{\lambda}}_n|\bm{\lambda}^0)$&$d(\hat{\bm{\lambda}}_n,\bm{\lambda}^0)$\\\cmidrule(lr){1-1}\cmidrule(lr){2-2}\cmidrule(lr){4-5}\cmidrule(lr){7-9}
A1&$n=$200&Bias&$-$0.0083 &0.0153 &&0.0707 &0.2113 &0.3225 \\
&&MSE&0.0026 &0.0071 &&&&\\
&$n=$500&Bias&$-$0.0026 &0.0099 &&0.0316 &0.0385 &0.0599 \\
&&MSE&0.0003 &0.0013 &&&&\\
&$n=$800&Bias&$-$0.0016 &0.0087 &&0.0247 &0.0259 &0.0462 \\
&&MSE&0.0002 &0.0012 &&&&\\\cmidrule(lr){1-9}
A2&$n=$200&Bias&$-$0.0004 &$-$0.0045 &&0.0390 &0.0392 &0.0615 \\
&&MSE&0.0010 &0.0011 &&&&\\
&$n=$500&Bias&$-$0.0008 &$-$0.0017 &&0.0146 &0.0109 &0.0183 \\
&&MSE&0.0001 &0.0001 &&&&\\
&$n=$800&Bias&$-$0.0005 &$-$0.0013 &&0.0090 &0.0066 &0.0126 \\
&&MSE&0.0000 &0.0000 &&&&\\\cmidrule(lr){1-9}
A3&$n=$200&Bias&$-$0.0049 &0.0051 &&0.0432 &0.0640 &0.0954 \\
&&MSE&0.0007 &0.0026 &&&&\\
&$n=$500&Bias&$-$0.0008 &0.0000 &&0.0165 &0.0136 &0.0221 \\
&&MSE&0.0001 &0.0004 &&&&\\
&$n=$800&Bias&$-$0.0006 &0.0008 &&0.0124 &0.0087 &0.0160 \\
&&MSE&0.0000 &0.0001 &&&&\\\midrule
Model&Sample size&&\multicolumn{1}{c}{$\lambda_1$}&\multicolumn{1}{c}{$\lambda_2$}&\multicolumn{1}{c}{$\lambda_3$}&$\zeta(\bm{\lambda}^0|\hat{\bm{\lambda}}_n)$&$\zeta(\hat{\bm{\lambda}}_n|\bm{\lambda}^0)$&$d(\hat{\bm{\lambda}}_n,\bm{\lambda}^0)$\\
\cmidrule(lr){1-1}\cmidrule(lr){2-2}\cmidrule(lr){4-6}\cmidrule(lr){7-9}
B1&$n=$200&Bias&$-$0.0104 &0.0090 &$-$0.0071 &0.0504 &0.4133 &0.6998 \\
&&MSE&0.0019 &0.0016 &0.0027 &&&\\
&$n=$500&Bias&$-$0.0035 &0.0049 &$-$0.0020 &0.0270 &0.1834 &0.3070 \\
&&MSE&0.0003 &0.0005 &0.0007 &&&\\
&$n=$800&Bias&$-$0.0018 &0.0022 &$-$0.0002 &0.0179 &0.0602 &0.0895 \\
&&MSE&0.0001 &0.0002 &0.0002 &&&\\\cmidrule(lr){1-9}
B2&$n=$200&Bias&0.0003 &$-$0.0021 &$-$0.0039 &0.0448 &0.0994 &0.1436 \\
&&MSE&0.0011 &0.0018 &0.0012 &&&\\
&$n=$500&Bias&0.0007 &$-$0.0006 &$-$0.0019 &0.0192 &0.0211 &0.0329 \\
&&MSE&0.0002 &0.0003 &0.0002 &&&\\
&$n=$800&Bias&0.0000 &$-$0.0002 &$-$0.0016 &0.0107 &0.0113 &0.0165 \\
&&MSE&0.0000 &0.0001 &0.0001 &&&\\\cmidrule(lr){1-9}
B3&$n=$200&Bias&0.0049 &$-$0.0011 &$-$0.0068 &0.0433 &0.1219 &0.1708 \\
&&MSE&0.0029 &0.0011 &0.0012 &&&\\
&$n=$500&Bias&0.0016 &$-$0.0002 &$-$0.0031 &0.0206 &0.0428 &0.0569 \\
&&MSE&0.0006 &0.0001 &0.0002 &&&\\
&$n=$800&Bias&$-$0.0002 &$-$0.0008 &$-$0.0016 &0.0125 &0.0203 &0.0259 \\
&&MSE&0.0002 &0.0000 &0.0000 &&&\\
\bottomrule
	\end{tabular}
\end{adjustwidth}
\end{table}

\begin{landscape}
\begin{table}
\scriptsize
\centering
\caption{Comparision of parameter estimation when the change-points are known and unknown.}\label{parameter estimate_Estimate_tau}
\begin{adjustwidth}{2cm}{1cm}
\begin{tabular}{ccc*{12}{r}}
\toprule
&&&\multicolumn{6}{c}{known change-points}&\multicolumn{6}{c}{unknown change-points}\\\cmidrule(lr){3-9}\cmidrule(lr){10-15}
Model&Sample size&&\multicolumn{1}{c}{$\rho_1$}&\multicolumn{1}{c}{$\rho_2$}&\multicolumn{1}{c}{$\rho_3$}&\multicolumn{1}{c}{$p_1$}&\multicolumn{1}{c}{$p_2$}&\multicolumn{1}{c}{$p_3$}&\multicolumn{1}{c}{$\rho_1$}&\multicolumn{1}{c}{$\rho_2$}&\multicolumn{1}{c}{$\rho_3$}&\multicolumn{1}{c}{$p_1$}&\multicolumn{1}{c}{$p_2$}&\multicolumn{1}{c}{$p_3$}\\\cmidrule(lr){1-1}\cmidrule(lr){2-2}\cmidrule(lr){4-6}\cmidrule(lr){7-9}\cmidrule(lr){10-12}\cmidrule(lr){13-15}
A1&$n=$200&Bias&$-$0.0175&$-$0.0004&$-$0.0679&0.0004 &0.0006 &0.0017 &$-$0.0273&$-$0.0162&$-$0.0901&0.0017 &0.0000 &0.0006 \\
&&MSE&0.0143 &0.0062 &0.0205 &0.0003 &0.0017 &0.0005 &0.0156 &0.0186 &0.0273 &0.0003 &0.0019 &0.0010 \\
&$n=$500&Bias&$-$0.0008&$-$0.0175&$-$0.0109&0.0006 &$-$0.0001&$-$0.0007&$-$0.0061&$-$0.0213&$-$0.0211&0.0008 &0.0000 &$-$0.0001\\
&&MSE&0.0053 &0.0038 &0.0070 &0.0001 &0.0005 &0.0002 &0.0058 &0.0039 &0.0079 &0.0001 &0.0006 &0.0003 \\
&$n=$800&Bias&$-$0.0039&$-$0.0094&0.0010 &0.0009 &0.0003 &$-$0.0009&$-$0.0057&$-$0.0098&$-$0.0042&0.0010 &$-$0.0003&$-$0.0007\\
&&MSE&0.0024 &0.0039 &0.0029 &0.0001 &0.0007 &0.0001 &0.0027 &0.0051 &0.0038 &0.0001 &0.0008 &0.0001 \\\cmidrule(lr){1-15}
A2&$n=$200&Bias&$-$0.0240&$-$0.0291&$-$0.0223&0.0001 &$-$0.0006&$-$0.0008&$-$0.0205&$-$0.0358&$-$0.0170&0.0006 &0.0002 &$-$0.0012\\
&&MSE&0.0136 &0.0166 &0.0154 &0.0004 &0.0005 &0.0005 &0.0148 &0.0214 &0.0189 &0.0005 &0.0007 &0.0006 \\
&$n=$500&Bias&$-$0.0106&$-$0.0065&$-$0.0101&$-$0.0006&$-$0.0009&0.0019 &$-$0.0110&$-$0.0072&$-$0.0094&$-$0.0001&$-$0.0008&0.0019 \\
&&MSE&0.0072 &0.0051 &0.0056 &0.0002 &0.0002 &0.0003 &0.0079 &0.0052 &0.0060 &0.0003 &0.0002 &0.0003 \\
&$n=$800&Bias&$-$0.0079&$-$0.0183&$-$0.0011&$-$0.0006&$-$0.0006&$-$0.0008&$-$0.0090&$-$0.0173&$-$0.0006&$-$0.0008&$-$0.0005&$-$0.0007\\
&&MSE&0.0028 &0.0056 &0.0026 &0.0001 &0.0003 &0.0001 &0.0031 &0.0060 &0.0027 &0.0001 &0.0003 &0.0001 \\\cmidrule(lr){1-15}
A3&$n=$200&Bias&0.0023 &$-$0.0329&$-$0.0280&$-$0.0001&$-$0.0040&0.0011 &$-$0.0019&$-$0.0249&$-$0.0295&0.0004 &$-$0.0080&0.0016 \\
&&MSE&0.0120 &0.0113 &0.0168 &0.0002 &0.0011 &0.0006 &0.0125 &0.0115 &0.0201 &0.0002 &0.0017 &0.0007 \\
&$n=$500&Bias&$-$0.0009&$-$0.0200&$-$0.0150&$-$0.0007&$-$0.0005&$-$0.0006&$-$0.0026&$-$0.0188&$-$0.0150&$-$0.0006&$-$0.0009&$-$0.0005\\
&&MSE&0.0072 &0.0033 &0.0070 &0.0001 &0.0006 &0.0002 &0.0072 &0.0033 &0.0073 &0.0001 &0.0007 &0.0003 \\
&$n=$800&Bias&0.0003 &$-$0.0215&$-$0.0020&0.0006 &0.0017 &0.0003 &$-$0.0002&$-$0.0172&$-$0.0015&0.0008 &0.0014 &0.0002 \\
&&MSE&0.0033 &0.0045 &0.0027 &0.0000 &0.0008 &0.0001 &0.0035 &0.0038 &0.0027 &0.0000 &0.0009 &0.0001 \\
\bottomrule
\end{tabular}
\end{adjustwidth}
\end{table}
\end{landscape}

\section{Real data analysis}

\begin{figure}
\begin{adjustwidth}{-2.5cm}{-0.5cm}
\centering
\includegraphics[scale=0.55]{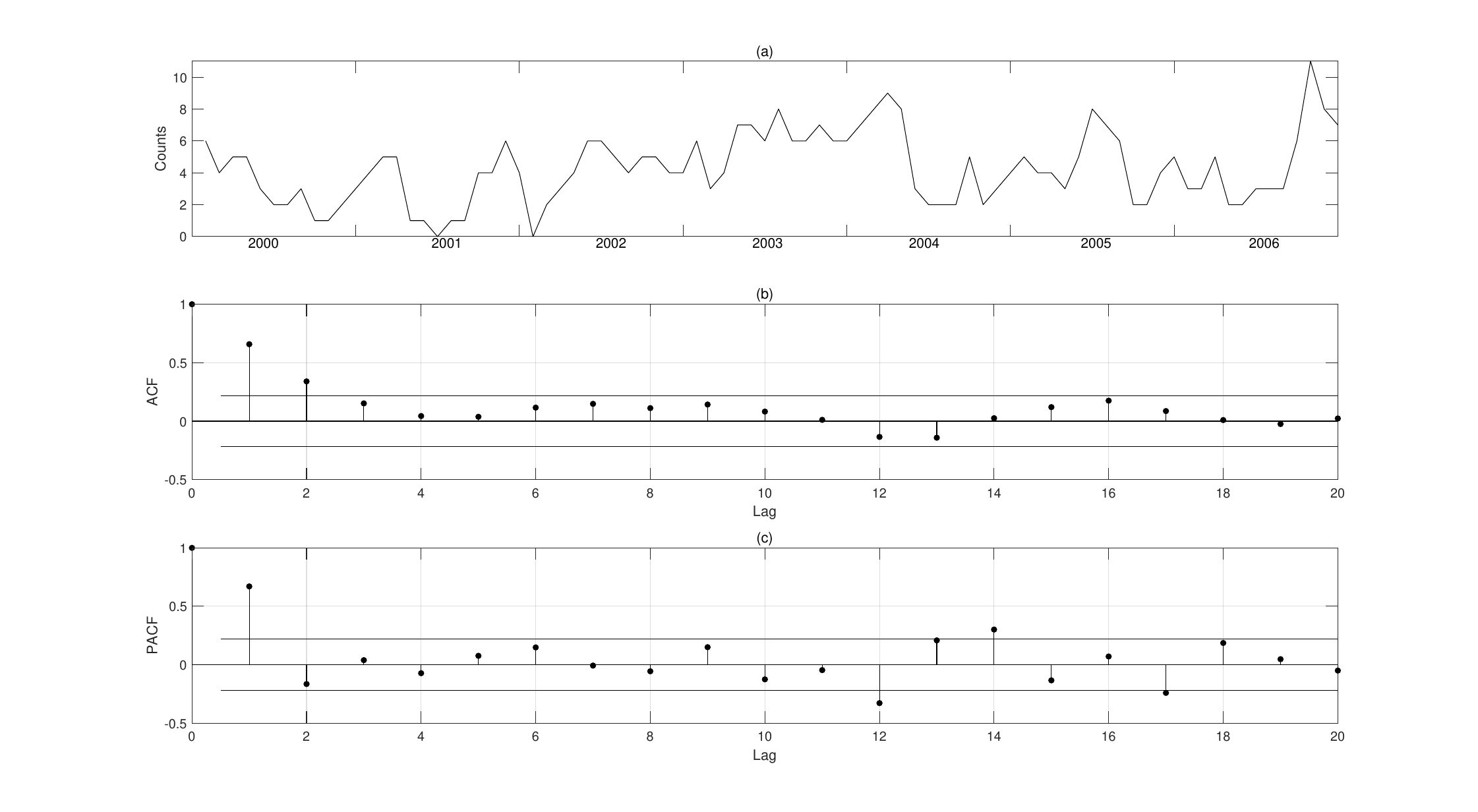}
\end{adjustwidth}
\begin{adjustwidth}{2cm}{1.5cm}
\caption{The sample path, ACF and PACF of the download counts \protect\\(from January 2000 to December 2006, sample size $n=84$).}\label{F1}
\end{adjustwidth}
\end{figure}

\begin{figure}
\begin{adjustwidth}{-2.5cm}{-0.5cm}
\centering
\includegraphics[scale=0.55]{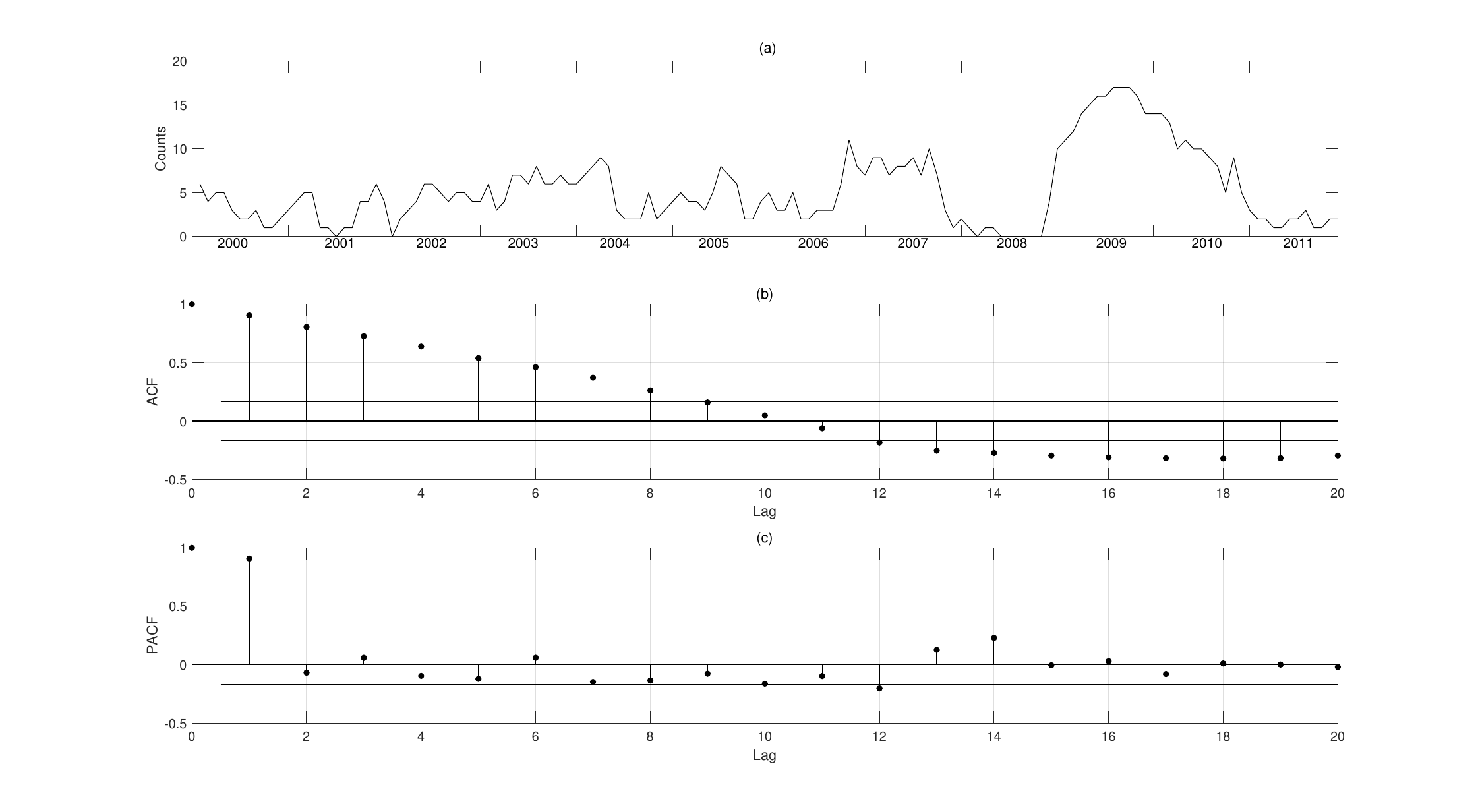}
\end{adjustwidth}
\begin{adjustwidth}{2cm}{1.5cm}
\caption{The sample path, ACF and PACF of the download counts\protect\\(from January 2000 to December 2011, sample size $n=143$).}\label{F2}
\end{adjustwidth}
\end{figure}

In this section, we conduct an application to demonstrate the usefulness of the MCP-BAR(1) model in explaining piecewise stationary phenomena in count time series with bounded support.
We applied the proposed model to fit the count data, which represent the number of seventeen European Union countries with inflation rates of less than 2$\%$ per month from January 2000 to December 2011.
This data set has been investigated by \cite{WeissKim2014}.
Especially, the authors focused on the data set during January 2000 to December 2006 and pointed out that the observations were quite stable in the years before 2007, but they became clear that the stationary behavior ends within 2007.
Furthermore, the authors found external evidence that supported their conjecture about a change-point within 2007.
The above conclusions can also be strongly supported by the time series plot, ACF, and PACF of the counts in Figures \ref{F1}-\ref{F2}.
Figure \ref{F1} shows that the observations during January 2000 to December 2006 are clearly stationary.
However, from Figure \ref{F2}, we can see that the ACF plot appears the shape of a symmetrical triangle and the time series plot also shows several significant trends and change-points after 2006.
These phenomena indicate that the data set is non-stationary.
In our work, we adopt the CUSUM test to explore whether it is necessary to use a BAR(1) model with change-points to fit this data set, i.e.,
the CUSUM test is used to solve the following testing problem:
$$\mbox{$\mathcal{H}_0$: $\rho$ and $p$ does not change over $X_1,...,X_{n}$ v.s. $\mathcal{H}_1$: not $\mathcal{H}_0$.}$$
The test statistics were computed as 23.9294 and 20.2223 based on the CLS and MQL estimators, respectively, which means that we reject $\mathcal{H}_0$ but accept $\mathcal{H}_1$ at the significance levels $\beta=0.01,0.05$ since the critical values are 3.269 and 2.408.
Hence, it is reasonable to employ the MCP-BAR(1) model to fit this data set.

For comparison purposes, we compare the MCP-BAR(1) model with
the BAR(1) model \citep{McKenzie1985}, Beta-BAR(1) model \citep{WeissKim2014}, and NBAR(1) model \citep{Zhang2023}.
It is worth mentioning that the NBAR(1) model is the BAR(1) model with one change-point.
The following statistics were employed to evaluate the capability of the fitted models: Akaike information criterion (AIC), Bayesian information criterion (BIC), root mean square
of differences between observations and forecasts (RMS), where RMS is calculated by
$${\rm RMS}=\sum_{j=1}^{\hat{m}}\sqrt{\frac{1}{n_j-1}\sum\limits_{t=2}^{n_j}\left[X_t-\hat{\rho}_j X_{t-1}-N\hat{p}_j(1-\hat{\rho}_j)\right]^2}.$$
From Table \ref{real_data}, we find that the BAR(1) model is not suitable to fit this data set since it gives poor performances based on each statistic.
Comparing the Beta-BAR(1) model with the NBAR(1) model, it is unexpected that the Beta-BAR(1) model outperforms the NBAR(1) model based on AIC and BIC since the NBAR(1) model can capture the change-point in observations.
The location of change-point in NBAR(1) model is estimated as $\hat{\tau}=107$, which is not consistent with the time series plot in Figure \ref{F2}.
For the MCP-BAR(1) model, it gives best performances among the alternative models based on each statistics.
Furthermore, we can see that the locations of change-points are $\hat{\tau}_{1}=91,\hat{\tau}_{2}=107$, and $\hat{\tau}_{3}=126$, which means that the change-points are observed during 2007, 2008 and 2010.
The estimation results are consistent with the time series plot in Figure \ref{F2}.


\begin{table}[H]
\centering
\caption{Estimators and statistics for price stability counts.}\label{real_data}
\begin{tabular}{*{9}{c}}
\toprule
Model&Segment&$\hat{\rho}$&$\hat{p}$&$\hat{\phi}$&$\hat{\tau}$&AIC&BIC&RMS\\\cmidrule(lr){1-1}\cmidrule(lr){2-2}\cmidrule(lr){3-6}\cmidrule(lr){7-9}
BAR&-&0.7470 &0.3079 &-&-&625.5306 &631.4563 &1.8814 \\\cmidrule(lr){1-9}
Beta-BAR(1)&-&0.3241 &0.7354 &0.1369 &-&569.5953 &578.4838 &1.9018 \\\cmidrule(lr){1-9}
NBAR(1)&I&0.6359 &0.3197 &-&107&581.0453 &592.8967 &1.6940 \\
&II&0.8478 &0.7784 &-&&&&\\\cmidrule(lr){1-9}
MCP-BAR(1)&I&0.6140 &0.2690 &-&91&518.2197 &541.9225 &1.6413 \\
&II&0.4464 &0.0595 &-&107&&&\\
&III&0.7635 &0.8072 &-&126&&&\\
&IV&0.5804 &0.1446 &-&&&&\\
\bottomrule
\end{tabular}
\end{table}

\section{Conclusion}
The target of this article is to introduce a new BAR(1) model with multiple change-points, which is useful to handle the non-stationary count time series with a finite range.
To detect the change-points, the CUSUM test is studied based on the CLS and MQL estimators.
The simulation studies show that the test statistics are effective to detect change-points.
Estimation for the number and locations of change-points is another important issue.
For this, the MDL principle and a new algorithm named S-GA are applied.
The simulation studies reveal that the adopted methods have the ability to give accurate estimators and save plenty of computing costs.
Finally, an application to the price stability counts is conducted to show the superiority of the multiple change-points BAR(1) model.

\section*{Acknowledgements}
The authors thank the associate editor and three reviewers for helpful comments, which led to a much improved version of the paper.
This paper is supported by National Natural Science Foundation of China (No. 12101485), China Postdoctoral Science Foundation (No. 2021M702624),
Fundamental Research Funds for the Central Universities (No. xzy012024037).
\section*{Appendix}
$\mathbf{Proof~of~Theorem~\ref{clsnormal}.}$
Note that the fact that the BAR(1) model is a strictly stationary and ergodic Markov chain and is bounded with all moments finite.
Thus, it is enough to check the assumption (B2) in Theorem 3.2.24 in \cite{Taniguchi2012}.
Let $m(\rho,p)={\rm E}(X_t|\mathcal{F}_{t-1})=\rho X_{t-1}+Np(1-\rho)$.
It follows that
\begin{align*}
\frac{\partial m(\rho,p)}{\partial\rho}=X_{t-1}-Np,~~~\frac{\partial m(\rho,p)}{\partial\rho}=N(1-\rho),
\end{align*}
which shows that the partial derivatives of the mean function form a linearly independent system.
After replacing $R$ by $\bm{W}_{CLS}$, $U$ by $\bm{V}_{CLS}$ in Theorem 3.2.24 in \cite{Taniguchi2012},
Theorem \ref{clsnormal} is easily to be proved.
~\\
~\\
$\mathbf{Proof~of~Theorem~\ref{test1}.}$
Since $\hat{\bm{\theta}}_{n,CLS}$ satisfies the least-squares equation, by Taylor's theorem, we obtain
\begin{align}
	\bm{0}=\frac{1}{\sqrt{n}}\frac{\partial S_{n}(\hat{\bm{\theta}}_{n,CLS})}{\partial \bm{\theta}}
	=\frac{1}{\sqrt{n}}\frac{\partial S_{n}(\bm{\theta}_0)}{\partial\bm{\theta}}+\frac{1}{n}\frac{\partial^2 S_{n}(\bm{\theta}_{n,CLS}^{*})}{\partial\bm{\theta}\partial\bm{\theta}^{\top}}\sqrt{n}(\hat{\bm{\theta}}_{n,CLS}-\bm{\theta}_0),\label{test_taylor}
\end{align}
where $\bm{\theta}_{n,CLS}^{*}$ is an intermediate point between $\bm{\theta}_0$ and $\hat{\bm{\theta}}_{n,CLS}$.
Denote
$$\bm{V}_{n,CLS}:=\frac{1}{2n}\frac{\partial^{2} S_{n}(\bm{\theta}_{n,CLS}^{*})}{\partial\bm{\theta}\partial\bm{\theta}^{\top}}.$$
By the ergodicity of $\{X_t\}$, it can be obtained that $\bm{V}_{n,CLS}\overset{\mathrm{a.s.}}{\longrightarrow} \bm{V}_{CLS}$ by using the ergodic theorem.
From Equation (\ref{test_taylor}), we have
\begin{align}\label{V_ninverse}
\bm{V}_{CLS}\cdot\sqrt{n}(\hat{\bm{\theta}}_{n,CLS}-\bm{\theta}_0)=-\frac{1}{2\sqrt{n}}\frac{\partial S_{n}(\bm{\theta}_0)}{\partial\bm{\theta}}+\left\{\bm{V}_{CLS}-\bm{V}_{n,CLS}\right\}\cdot\sqrt{n}(\hat{\bm{\theta}}_{n,CLS}-\bm{\theta}_0).
\end{align}
Furthermore, if the inverse matrix of $\bm{V}_{n,CLS}$ exists, we have
\begin{align*}
\sqrt{n}(\hat{\bm{\theta}}_{n,CLS}-\bm{\theta}_0)=-\left(\bm{V}_{n,CLS}\right)^{-1}\cdot\frac{1}{2\sqrt{n}}\frac{\partial S_{n}(\bm{\theta}_0)}{\partial\bm{\theta}},
\end{align*}
then
\begin{align}\label{V_inverse}
\bm{V}_{CLS}\cdot\sqrt{n}(\hat{\bm{\theta}}_{n,CLS}-\bm{\theta}_0)=&-\frac{1}{2\sqrt{n}}\frac{\partial S_{n}(\bm{\theta}_0)}{\partial\bm{\theta}}\nonumber\\
&-\left(\bm{V}_{CLS}-\bm{V}_{n,CLS}\right)\cdot\left(\bm{V}_{n,CLS}\right)^{-1}\cdot\frac{1}{2\sqrt{n}}\frac{\partial S_{n}(\bm{\theta}_0)}{\partial\bm{\theta}}.
\end{align}
According to Equations (\ref{V_ninverse}) and (\ref{V_inverse}), we can rewrite that for $0\leqslant \lambda\leqslant1$,
\begin{align}\label{Appendix.1}
\bm{V}_{CLS}\cdot\frac{[n\lambda]}{\sqrt{n}}(\hat{\bm{\theta}}_{[n\lambda],CLS}-\hat{\bm{\theta}}_{n,CLS})=&-\frac{1}{2\sqrt{n}}\frac{\partial S_{[n\lambda]}(\bm{\theta}_0)}{\partial\bm{\theta}}-\frac{[n\lambda]}{n}\left(-\frac{1}{2\sqrt{n}}\frac{\partial S_{n}(\bm{\theta}_0)}{\partial\bm{\theta}}\right)\nonumber\\
&+\sqrt{\frac{[n\lambda]}{n}}\Delta_{[n\lambda]}-\frac{[n\lambda]}{n}\Delta_{n},
\end{align}
where
\begin{align*}
\Delta_{k}=
\begin{cases}	
-\left(\bm{V}_{CLS}-\bm{V}_{k,CLS}\right)\cdot\left(\bm{V}_{k,CLS}\right)^{-1}\cdot\frac{1}{2\sqrt{k}}\frac{\partial S_{k}(\bm{\theta}_0)}{\partial\bm{\theta}},~\text{if}~\bm{V}_{k,CLS}^{-1}~\text{exists},\\
\left(\bm{V}_{CLS}-\bm{V}_{k,CLS}\right)\cdot\sqrt{k}(\hat{\bm{\theta}}_{k,CLS}-\bm{\theta}_0),~\text{otherwise.}
\end{cases}
\end{align*}
It is easy to check that $\mathrm{E}[\partial S_{k}(\bm{\theta}_0)/\partial\bm{\theta}|\mathcal{F}_{k-1}]={\bm 0}$.
As seen in \cite{Lee2003}, the functional limit theorem for martingales is a key tool to verify the asymptotic results for the CUSUM
test, coupled with the fact that $\bm{W}_{CLS}$ is a positive definite matrix under $\mathcal{H}_0$, there is
%
\begin{align}\label{Appendix.2}
-\bm{W}_{CLS}^{-1/2}\frac{1}{2\sqrt{n}}\frac{\partial S_{[n\lambda]}(\bm{\theta}_0)}{\partial\bm{\theta}}\mathop{\longrightarrow}\limits^{\mathrm{d}}\mathbf{B}_2^{\circ}(\lambda),
\end{align}
where $\mathbf{B}_2^{\circ}(\lambda)=(B_1^{\circ}(\lambda),B_2^{\circ}(\lambda))^{\top}$ is a two-dimensional standard Brownian motion.
Building upon Lemma \ref{lemma1}, which asserts that
$$\max_{1\leqslant k\leqslant n}\sqrt{\frac{k}{n}}||\Delta_{k}||=o_{p}(1),$$
and integrating the results from (\ref{Appendix.1}) and (\ref{Appendix.2}), we have
$$\bm{W}_{CLS}^{-1/2}\bm{V}_{CLS}\frac{[n\lambda]}{\sqrt{n}}(\hat{\bm{\theta}}_{[n\lambda],CLS}-\hat{\bm{\theta}}_{n,CLS}) \mathop{\longrightarrow}\limits^{\mathrm{d}}\bm{B}_2(\lambda).$$
The proof of Theorem \ref{test1} has been completed.
~\\
~\\
$\mathbf{Proof~of~Theorem~\ref{the_cls_c}.}$
Following Theorem \ref{test1}, the result in Theorem \ref{the_cls_c} under $\mathcal{H}_0$ is obviously true. Inspired by the proof of Theorem 2 in \cite{Pesta2020},	
we next prove the result in Theorem \ref{the_cls_c} under $\mathcal{H}_1$, i.e,
$$C_n^{CLS}\mathop{\longrightarrow}\limits^{\mathrm{p}}+\infty,~~n\rightarrow\infty.$$
Without loss of generality, we assume that the sequence $\{X_t\}_{t=1}^n$ has only one change point at $t=\tau$, and denote $\lambda=\tau/n$ with $0<\lambda<1$.
It follows that we can divide the sequence $\{X_t\}_{t=1}^n$ into two stationary sequences, $\{X_t\}_{t=1}^{\tau}$ and $\{X_t\}_{t=\tau+1}^n$, which are generated from  the BAR(1) model (\ref{BAR(1)}) depending on $\bm{\theta}_1$ and $\bm{\theta}_2$  with $\bm{\theta}_1\neq\bm{\theta}_2$, respectively.
According to the definition of $C_n^{CLS}$, we have
\begin{align*}
C_n^{CLS}\geqslant C_{n,\tau}^{CLS}=\frac{\tau^2}{n}(\hat{\bm{\theta}}_{\tau,CLS}-\hat{\bm{\theta}}_{n,CLS})^{\top}\hat{\bm{V}}_{n,CLS}\hat{\bm{W}}_{n,CLS}^{-1}\hat{\bm{V}}_{n,CLS}(\hat{\bm{\theta}}_{\tau,CLS}-\hat{\bm{\theta}}_{n,CLS}).
\end{align*}
Following the asymptotic proprieties of the CLS estimator, we have $\hat{\bm{\theta}}_{\tau,CLS}\mathop{\longrightarrow}\limits^{\mathrm{a.s.}}\bm{\theta}_1$,
as $n\rightarrow\infty$.
Also, since $\bm{\theta}_1\neq\bm{\theta}_2$, the asymptotic proprieties imply that
$$||\hat{\bm{\theta}}_{\tau,CLS}-\hat{\bm{\theta}}_{n,CLS}||\neq 0,~~n\rightarrow\infty.$$
Notably, the determinant of $\hat{\bm{V}}_{n,CLS}$ is given by
\begin{align*}
|\hat{\bm{V}}_{n,CLS}|&=N^2(1-\hat{\rho}_{n,CLS})^2\frac{1}{n}\sum_{t=1}^n(X_{t-1}-N\hat{p}_{n,CLS})^2-
\left[\frac{1}{n}\sum_{t=1}^{n}(X_{t-1}-N\hat{p}_{n,CLS})N(1-\hat{\rho}_{n,CLS})\right]^2\\
&=N^2(1-\hat{\rho}_{n,CLS})^2\left\{\frac{1}{n}\sum_{t=1}^n(X_{t-1}-N\hat{p}_{n,CLS})^2-\left[\frac{1}{n}\sum_{t=1}^{n}(X_{t-1}-N\hat{p}_{n,CLS})\right]^2\right\}.
\end{align*}
Clearly, $|\hat{\bm{V}}_{n,CLS}|\neq 0$. Similarly, we can verify $|\hat{\bm{W}}_{n,CLS}|\neq 0$. Thus, there exist a constant $c\neq0$ such that
\begin{align*}
C_n^{CLS}&\geqslant C_{n,\tau}^{CLS}
=\lambda([n\lambda])(\hat{\bm{\theta}}_{\tau,CLS}-\hat{\bm{\theta}}_{n,CLS})^{\top}\hat{\bm{V}}_{n,CLS}\hat{\bm{W}}_{n,CLS}^{-1}\hat{\bm{V}}_{n,CLS}(\hat{\bm{\theta}}_{\tau,CLS}-\hat{\bm{\theta}}_{n,CLS})\\
&\geqslant c\lambda[n\lambda]\mathop{\longrightarrow}\limits^{\mathrm{p}}+\infty,~~n\rightarrow\infty.
\end{align*}
That is, $C_n^{CLS}\mathop{\longrightarrow}\limits^{\mathrm{p}}+\infty,$ as $n\rightarrow\infty$. The proof of Theorem \ref{the_cls_c} has been completed.\\
~\\
$\mathbf{Proof~of~Theorem~\ref{the_qml_c}.}$ The proof is similar to the proof of Theorem \ref{the_cls_c}, and we omit it.\\
~\\
$\mathbf{Expression~in~Theorem~\ref{the_qml_c}.}$
Let $k_0$ be a positive integer, and define
\begin{align*}
C_{n}^{MQL}=\max_{k_0\leqslant k\leqslant n}\frac{k^2}{n}(\hat{\bm{\theta}}_{k,MQL}-\hat{\bm{\theta}}_{n,MQL})^{\top}\hat{\bm{V}}_{n,MQL}\hat{\bm{W}}_{n,MQL}^{-1}\hat{\bm{V}}_{n,MQL}(\hat{\bm{\theta}}_{k,MQL}-\hat{\bm{\theta}}_{n,MQL}).
\end{align*}
where
\begin{align*}\small
\hat{\bm{V}}_{n,MQL}=\left({\begin{array}{lc}
\hat{{V}}_{n,MQL}^{(1,1)}&\hat{{V}}_{n,MQL}^{(1,2)}\\
\hat{{V}}_{n,MQL}^{(2,1)}&\hat{{V}}_{n,MQL}^{(2,2)}
\end{array}}\right),~~~
\hat{\bm{W}}_{n,MQL}=\left( {\begin{array}{*{2}{c}}
\hat{{W}}_{n,MQL}^{(1,1)}&\hat{{W}}_{n,MQL}^{(1,2)}\\
\hat{{W}}_{n,MQL}^{(2,1)}&\hat{{W}}_{n,MQL}^{(2,2)}
\end{array}} \right)
\end{align*}
with
\begin{align*}
&\hat{{V}}_{n,MQL}^{(1,1)}={\frac{1}{n}\sum\limits_{t=1}^nD_{t}(\hat{\bm{\theta}}_{n,CLS})(X_{t-1}-N\hat{p}_{n,MQL})^2},\\
&\hat{{V}}_{n,MQL}^{(1,2)}=\hat{{V}}_{n,MQL}^{(2,1)}={\frac{1}{n}\sum\limits_{t=1}^{n}D_{t}(\hat{\bm{\theta}}_{n,CLS})\big{\{}(X_{t-1}-N\hat{p}_{n,MQL})(N(1-\hat{\rho}_{n,MQL})+s_t(\bm{\theta}_{n,MQL})]\big{\}}},\\
&\hat{{V}}_{n,MQL}^{(2,2)}={N^2(1-\hat{\rho}_{n,MQL})^2\frac{1}{n}\sum\limits_{t=1}^{n}D_{t}(\hat{\bm{\theta}}_{n,CLS})},\\
&\hat{{W}}_{n,MQL}^{(1,1)}={\frac{1}{n}\sum_{t=1}^{n}[D_{t}(\hat{\bm{\theta}}_{n,CLS})s_t(\hat{\bm{\theta}}_{n,MQL})(X_{t-1}-N\hat{p}_{n,MQL})]^2},\\
&\hat{{W}}_{n,MQL}^{(1,2)}=\hat{{W}}_{n,MQL}^{(2,1)}={\frac{1}{n}\sum_{t=1}^{n}[D_{t}(\hat{\bm{\theta}}_{n,CLS})s_t(\hat{\bm{\theta}}_{n,MQL})]^2(X_{t-1}-N\hat{p}_{n,MQL})N(1-\hat{\rho}_{n,MQL})},\\
&\hat{{W}}_{n,MQL}^{(2,2)}={\frac{1}{n}\sum_{t=1}^{n}[D_{t}(\hat{\bm{\theta}}_{n,CLS})s_t(\hat{\bm{\theta}}_{n,MQL})N(1-\hat{\rho}_{n,MQL})]^2}.
\end{align*}
~\\
$\mathbf{Proof~of~Theorem~\ref{the_cml_c}.}$ The proof is similar to the proof of Theorem \ref{the_cls_c}, and we omit it.\\
~\\
$\mathbf{Fisher~information~matrix~Expression~in~Theorem~\ref{the_cml_c}.}$
Recall that the transition probabilities of the BAR(1) model is given by
\begin{align*}
\mathrm{P}(X_t|X_{t-1})
 &=\sum_{k=\max\{0,X_{t-1}+X_t-N\}}^{\min\{X_{t-1},X_t\}}\binom{X_{t-1}}{k}\binom{N-X_{t-1}}{X_t-k}\alpha^{k}(1-\alpha)^{X_{t-1}-k}\beta^{X_t-k}(1-\beta)^{N-X_{t-1}-X_t+k},\\
 &=:\sum_{k=\Delta_1}^{\Delta_2} J_0(k)J_1(k,\bm{\theta})J_2(k,\bm{\theta})J_3(k,\bm{\theta})J_4(k,\bm{\theta}),
\end{align*}
where
\begin{align*}
&J_0(k)=\binom{X_{t-1}}{k}\binom{N-X_{t-1}}{X_t-k},\Delta_1=\max\{0,X_{t-1}+X_t-N\}, \Delta_2=\min\{X_{t-1},X_t\},\\
&J_1(k,\bm{\theta})=\alpha^{k},J_2(k,\bm{\theta})=(1-\alpha)^{X_{t-1}-k}, J_3(k,\bm{\theta})=\beta^{X_t-k},J_4(k,\bm{\theta})=(1-\beta)^{N-X_{t-1}-X_t+k}.
\end{align*}
It follows that
\begin{align*}
&\frac{\partial J_1(k,\bm{\theta})}{\partial\rho}=k\alpha^{k-1}(1-p),~~~\frac{\partial J_1(k,\bm{\theta})}{\partial p}=k\alpha^{k-1}(1-\rho),\\
&\frac{\partial J_2(k,\bm{\theta})}{\partial\rho}=-(X_{t-1}-k)(1-\alpha)^{X_{t-1}-k-1}(1-p),\\
&\frac{\partial J_2(k,\bm{\theta})}{\partial p}=-(X_{t-1}-k)(1-\alpha)^{X_{t-1}-k-1}(1-\rho),\\
&\frac{\partial J_3(k,\bm{\theta})}{\partial\rho}=({X_t-k})\beta^{X_t-k-1}(-p),~~~\frac{\partial J_3(k,\bm{\theta})}{\partial p}=({X_t-k})\beta^{X_t-k-1}(1-\rho),\\
&\frac{\partial J_4(k,\bm{\theta})}{\partial\rho}=-({N-X_{t-1}-X_t+k})(1-\beta)^{N-X_{t-1}-X_t+k-1}(-p),\\
&\frac{\partial J_4(k,\bm{\theta})}{\partial p}=-({N-X_{t-1}-X_t+k})(1-\beta)^{N-X_{t-1}-X_t+k-1}(1-\rho).
\end{align*}
After a simple calculation, there is
\begin{align*}
{\rm P}_{\rho}^{'}(k,\bm{\theta})=&\frac{\partial \mathrm{P}(X_t|X_{t-1})}{\partial \rho}=\sum_{k=\Delta_1}^{\Delta_2} J_0(k)\Big[
\frac{\partial J_1(k,\bm{\theta})}{\partial \rho}J_2(k,\bm{\theta})J_3(k,\bm{\theta})J_4(k,\bm{\theta})\\
&+\frac{\partial J_2(k,\bm{\theta})}{\partial \rho}J_1(k,\bm{\theta})J_3(k,\bm{\theta})J_4(k,\bm{\theta})
+\frac{\partial J_3(k,\bm{\theta})}{\partial \rho}J_1(k,\bm{\theta})J_2(k,\bm{\theta})J_4(k,\bm{\theta})\\
&+\frac{\partial J_4(k,\bm{\theta})}{\partial \rho}J_1(k,\bm{\theta})J_2(k,\bm{\theta})J_3(k,\bm{\theta})\Big],\\
{\rm P}_{p}^{'}(k,\bm{\theta})=&\frac{\partial \mathrm{P}(X_t|X_{t-1})}{\partial p}=\sum_{k=\Delta_1}^{\Delta_2} J_0(k)\Big[
\frac{\partial J_1(k,\bm{\theta})}{\partial p}J_2(k,\bm{\theta})J_3(k,\bm{\theta})J_4(k,\bm{\theta})\\
&+\frac{\partial J_2(k,\bm{\theta})}{\partial p}J_1(k,\bm{\theta})J_3(k,\bm{\theta})J_4(k,\bm{\theta})
+\frac{\partial J_3(k,\bm{\theta})}{\partial p}J_1(k,\bm{\theta})J_2(k,\bm{\theta})J_4(k,\bm{\theta})\\
&+\frac{\partial J_4(k,\bm{\theta})}{\partial p}J_1(k,\bm{\theta})J_2(k,\bm{\theta})J_3(k,\bm{\theta})\Big].
\end{align*}
Furthermore, there is
\begin{align*}
&\frac{\partial^2 J_1(k,\bm{\theta})}{\partial\rho^2}=k({k-1})\alpha^{k-2}(1-p)^2,~~\frac{\partial^2 J_1(k,\bm{\theta})}{\partial p^2}=k({k-1})\alpha^{k-2}(1-\rho)^2,\\
&\frac{\partial^2 J_1(k,\bm{\theta})}{\partial\rho\partial p}=\frac{\partial^2 J_1(k,\bm{\theta})}{\partial p\partial\rho}=
k({k-1})(\alpha)^{k-2}(1-p)(1-\rho)-k\alpha^{k-1},\\
&\frac{\partial^2 J_2(k,\bm{\theta})}{\partial\rho^2}=({X_{t-1}-k-1})(X_{t-1}-k)(1-\alpha)^{X_{t-1}-k-2}(1-p)^2,\\
&\frac{\partial^2 J_2(k,\bm{\theta})}{\partial p^2}=({X_{t-1}-k-1})(X_{t-1}-k)(1-\alpha)^{X_{t-1}-k-2}(1-\rho)^2,\\
&\frac{\partial^2 J_2(k,\bm{\theta})}{\partial\rho\partial p}=\frac{\partial^2 J_2(k,\bm{\theta})}{\partial p\partial\rho}=({X_{t-1}-k-1})(X_{t-1}-k)(1-\alpha)^{X_{t-1}-k-2}(1-p)(1-\rho)\\
&~~~~~~~~~~~~~~~~~~~~~~~~~~~~~~~~~~~+(X_{t-1}-k)(1-\alpha)^{X_{t-1}-k-1},\\
&\frac{\partial^2 J_3(k,\bm{\theta})}{\partial\rho^2}=({X_t-k-1})({X_t-k})\beta^{X_t-k-2}p^2,\\
&\frac{\partial^2 J_3(k,\bm{\theta})}{\partial p^2}=({X_t-k-1})({X_t-k})\beta^{X_t-k-2}(1-\rho)^2,\\
&\frac{\partial^2 J_3(k,\bm{\theta})}{\partial\rho\partial p}=\frac{\partial^2 J_3(k,\bm{\theta})}{\partial p\partial\rho}=({X_t-k-1})({X_t-k})\beta^{X_t-k-2}p(\rho-1)-({X_t-k})\beta^{X_t-k-1},\\
&\frac{\partial^2 J_4(k,\bm{\theta})}{\partial\rho^2}=({N-X_{t-1}-X_t+k-1})({N-X_{t-1}-X_t+k})(1-\beta)^{N-X_{t-1}-X_t+k-2}p^2,\\
&\frac{\partial^2 J_4(k,\bm{\theta})}{\partial p^2}=({N-X_{t-1}-X_t+k-1})({N-X_{t-1}-X_t+k})(1-\beta)^{N-X_{t-1}-X_t+k-2}(1-\rho)^2,\\
&\frac{\partial^2 J_4(k,\bm{\theta})}{\partial\rho\partial p}=\frac{\partial^2 J_4(k,\bm{\theta})}{\partial p\partial\rho}=({N-X_{t-1}-X_t+k-1})({N-X_{t-1}-X_t+k})(1-\beta)^{N-X_{t-1}-X_t+k-2}\\
&~~~~~~~~~~~~~~~~~~~~~~~~~~~~~~~~~~~\times p(\rho-1)+({N-X_{t-1}-X_t+k})(1-\beta)^{N-X_{t-1}-X_t+k-1}.
\end{align*}

Denote
\begin{align*}
A_{1j}(k,\bm{\theta})=&\frac{\partial^2 J_j(k,\bm{\theta})}{\partial\rho^2}\prod_{s_1=1, \hfill\atop s_1\neq j}^4J_{s_1}(k,\bm{\theta})+
\frac{\partial J_j(k,\bm{\theta})}{\partial\rho}\left[\sum_{s_2=1,\hfill\atop s_2\neq j}^4
\left(\frac{\partial J_{s_2}(k,\bm{\theta})}{\partial\rho}\prod_{s_3=1,\hfill s_3\neq s_2, \atop s_3\neq j}^4 J_{s_3}(k,\bm{\theta})\right)\right],
\end{align*}
\begin{align*}
A_{2j}(k,\bm{\theta})=&\frac{\partial^2 J_j(k,\bm{\theta})}{\partial\rho\partial p}
\prod_{s_1=1, \hfill\atop s_1\neq j}^4J_{s_1}(k,\bm{\theta})+
\frac{\partial J_j(k,\bm{\theta})}{\partial\rho}\left[\sum_{s_2=1,\hfill\atop s_2\neq j}^4
\left(\frac{\partial J_{s_2}(k,\bm{\theta})}{\partial p}\prod_{s_3=1,\hfill s_3\neq s_2, \atop s_3\neq j}^4 J_{s_3}(k,\bm{\theta})\right)\right],
\end{align*}
\begin{align*}
A_{3j}(k,\bm{\theta})=&\frac{\partial^2 J_j(k,\bm{\theta})}{\partial p^2}
\prod_{s_1=1, \hfill\atop s_1\neq j}^4J_{s_1}(k,\bm{\theta})+
\frac{\partial J_j(k,\bm{\theta})}{\partial p}\left[\sum_{s_2=1,\hfill\atop s_2\neq j}^4
\left(\frac{\partial J_{s_2}(k,\bm{\theta})}{\partial p}\prod_{s_3=1,\hfill s_3\neq s_2, \atop s_3\neq j}^4 J_{s_3}(k,\bm{\theta})\right)\right],
\end{align*}
where $j=1,2,3,4$.
Then, after some simple calculations, there is
\begin{align*}
&{\rm P}_{\rho}^{''}(k,\bm{\theta})=\frac{\partial^2 \mathrm{P}(X_t|X_{t-1})}{\partial \rho^2}=\sum_{k=\Delta_1}^{\Delta_2} J_0(k)\left[A_{11}(k,\bm{\theta})+A_{12}(k,\bm{\theta})+A_{13}(k,\bm{\theta})+A_{14}(k,\bm{\theta})\right],\\
&{\rm P}_{\rho,p}^{''}(k,\bm{\theta})=\frac{\partial^2 \mathrm{P}(X_t|X_{t-1})}{\partial \rho\partial p}=\frac{\partial^2 \mathrm{P}(X_t|X_{t-1})}{\partial p\partial\rho}\\
&~~~~~~~~~~~~=\sum_{k=\Delta_1}^{\Delta_2} J_0(k)\left[A_{21}(k,\bm{\theta})+A_{22}(k,\bm{\theta})+A_{23}(k,\bm{\theta})+A_{24}(k,\bm{\theta})\right],\\
&{\rm P}_{p}^{''}(k,\bm{\theta})=\frac{\partial^2 \mathrm{P}(X_t|X_{t-1})}{\partial p^2}=\sum_{k=\Delta_1}^{\Delta_2} J_0(k)\left[A_{31}(k,\bm{\theta})+A_{32}(k,\bm{\theta})+A_{33}(k,\bm{\theta})+A_{34}(k,\bm{\theta})\right].
\end{align*}
Recall that the Fisher information matrix is
$$\bm{I}^{-1}(\bm{\theta}_0)=\mathrm{E}\left[-\frac{\partial^2 \ell_t(\bm{\theta}_0|X_{t-1})}{\partial \bm{\theta}\partial \bm{\theta}^{\mathrm{T}}}\right].$$
Clearly, according to the ergodicity of the BAR(1) model, the consistent estimator of $\bm{I}^{-1}(\bm{\theta}_0)$ is given by
\begin{align*}
\hat{\bm{I}}_{n,CML}^{-1}=\left({\begin{array}{*{2}{c}}
\hat{I}_{n,CML}^{(1,1)}&\hat{I}_{n,CML}^{(1,2)}\\
\hat{I}_{n,CML}^{(1,2)}&\hat{I}_{n,CML}^{(2,2)}
\end{array}}\right),
\end{align*}
where
\begin{align*}
\hat{I}_{n,CML}^{(1,1)}&=\frac{1}{n}\sum_{t=1}^n\left\{-\frac{1}{\left(\mathrm{P}(X_t|X_{t-1})\right)^2}\left[{\rm P}_{\rho}^{'}(k,\bm{\theta})\right]^2
+\frac{1}{\mathrm{P}(X_t|X_{t-1})}{\rm P}_{\rho}^{''}(k,\bm{\theta})\right\}_{\bm{\theta}=\hat{\bm{\theta}}_{n,CML}},\\
\hat{I}_{n,CML}^{(1,2)}&=\frac{1}{n}\sum_{t=1}^n\left\{-\frac{1}{\left(\mathrm{P}(X_t|X_{t-1})\right)^2}\left[{\rm P}_{\rho}^{'}(k,\bm{\theta}){\rm P}_{p}^{'}(k,\bm{\theta})\right]
+\frac{1}{\mathrm{P}(X_t|X_{t-1})}{\rm P}_{\rho,p}^{''}(k,\bm{\theta})\right\}_{\bm{\theta}=\hat{\bm{\theta}}_{n,CML}},\\
\hat{I}_{n,CML}^{(2,2)}&=\frac{1}{n}\sum_{t=1}^n\left\{-\frac{1}{\left(\mathrm{P}(X_t|X_{t-1})\right)^2}\left[{\rm P}_{p}^{'}(k,\bm{\theta})\right]^2
+\frac{1}{\mathrm{P}(X_t|X_{t-1})}{\rm P}_{p}^{''}(k,\bm{\theta})\right\}_{\bm{\theta}=\hat{\bm{\theta}}_{n,CML}}.
\end{align*}

$\mathbf{Proof~of~Theorem~\ref{strong}.}$
Let the time series $\{X_{t,j}\}_{t=1}^{n_j}$ generate from the $j$th segment BAR model. Denote the true likelihood based on the time series $\{X_{t,j}\}_{t=1}^{n_j}$ by
\begin{align}\label{CMLfunction}
\widetilde{L}_{n_j}(0,\bm{\theta}_j)&=\sum\limits_{t=1}^{n_j}\ell_t(\bm{\theta}_j|X_{s,j},s<t)=\sum\limits_{t=1}^{n_j}\log\mathrm{P}(X_{t,j}|X_{t-1,j}).
\end{align}
Denote
$\sup_{\lambda_d,\lambda_u}:=\sup_{\lambda_d\in[0,1],\lambda_u\in[0,1],\lambda_u-\lambda_d>\epsilon_{\lambda}}.$
Define, for $j=1,...,m+1$, the true and the observed likelihood formed by a portion of the $j$th segment respectively by
\begin{align*}
\widetilde{L}_{n_j}(\bm{\theta}_j,\lambda_d,\lambda_u)&=\sum\limits_{t=[n_j\lambda_d]}^{[n_j\lambda_u]}\ell_t(\bm{\theta}_j|X_{s,j},s<t)=\sum\limits_{t=[n_j\lambda_d]}^{[n_j\lambda_u]}\log\mathrm{P}(X_{t,j}|X_{t-1,j}),\\
L_{n_j}(\bm{\theta}_j,\lambda_d,\lambda_u)&=\sum\limits_{t=[n_j\lambda_d]}^{[n_j\lambda_u]}\ell_t(\bm{\theta}_j|X_s,s<t)=\sum\limits_{t=[n_j\lambda_d]}^{[n_j\lambda_u]}\log\mathrm{P}(X_t|X_{t-1}).
\end{align*}
To ensure the validity of Theorem~\ref{strong}, we confirm adherence to Assumptions 1 (2), 2 (4), 3 , 5, and either 4 (0.5) or 4* in \cite{DavisYau2013}.
Given our focus on the first-order BAR model, Assumption 5 in \cite{DavisYau2013}, which is designed to ensure model selection consistency and identifiability of models, is obviously true.
For easy reading, we summarize Assumptions 1($\kappa$), 2($\kappa$), 3,  4* in \cite{DavisYau2013} as follows:

$\bullet$ Assumption 1($\kappa$): For any $j=1,2,...,m+1$, the function $\ell_t(\bm{\theta}_j|X_{s,j},s<t)$ is two-time continuously differentiable with respective to $\bm{\theta}_j$,
and the first and second derivatives $\widetilde{L}_{n_j}^{'}(\bm{\theta}_j,\lambda_d,\lambda_u)$, $L_{n_j}^{'}(\bm{\theta}_j,\lambda_d,\lambda_u)$ and $\widetilde{L}_{n_j}^{''}(\bm{\theta}_j,\lambda_d,\lambda_u)$, $L_{n_j}^{''}(\bm{\theta}_j,\lambda_d,\lambda_u)$, respectively, of the function $\widetilde{L}_{n_j}(\bm{\theta}_j,\lambda_d,\lambda_u)$ and $L_{n_j}(\bm{\theta}_j,\lambda_d,\lambda_u)$, satisfy
\begin{align*}
\sup_{\lambda_d,\lambda_u}\sup\limits_{\bm{\theta}_j\in\Theta_j}|\frac{1}{n}\widetilde{L}_{n_j}(\bm{\theta}_j,\lambda_d,\lambda_u)-\frac{1}{n}L_{n_j}(\bm{\theta}_j,\lambda_d,\lambda_u)|&=o(n^{\frac{1}{\kappa}-1}),\\
\sup_{\lambda_d,\lambda_u}\sup\limits_{\bm{\theta}_j\in\Theta_j}|\frac{1}{n}\widetilde{L}_{n_j}^{'}(\bm{\theta}_j,\lambda_d,\lambda_u)-\frac{1}{n}L_{n_j}^{'}(\bm{\theta}_j,\lambda_d,\lambda_u)|&=o(n^{\frac{1}{\kappa}-1}),\\
\sup_{\lambda_d,\lambda_u}\sup\limits_{\bm{\theta}_j\in\Theta_j}|\frac{1}{n}\widetilde{L}_{n_j}^{''}(\bm{\theta}_j,\lambda_d,\lambda_u)-\frac{1}{n}L_{n_j}^{''}(\bm{\theta}_j,\lambda_d,\lambda_u)|&=o(1),
\end{align*}
almost surely.

$\bullet$ Assumption 2($\kappa$): For $j=1,2,...,m+1$, there exists an $\epsilon>0$ such that
\begin{align*}
&\sup\limits_{\bm{\theta}_j\in\Theta_j}{\rm E}|\ell_t(\bm{\theta}_j|X_{s,j},s<t)|^{\kappa+\epsilon}<\infty,\\
&\sup\limits_{\bm{\theta}_j\in\Theta_j}{\rm E}|\ell_t^{'}(\bm{\theta}_j|X_{s,j},s<t)|^{\kappa+\epsilon}<\infty,\\
&\sup\limits_{\bm{\theta}_j\in\Theta_j}{\rm E}|\ell_t^{''}(\bm{\theta}_j|X_{s,j},s<t)|<\infty,
\end{align*}
where $\ell_t^{'}$ and $\ell_t^{''}$ are the first and second derivatives of $\ell_t$

$\bullet$ Assumption 3: For each $j=1,2,...,m+1$,
\begin{align*}
&\sup\limits_{\bm{\theta}_j\in\Theta_j}\left|\frac{1}{n}\widetilde{L}_{n_j}(0,\bm{\theta}_j)-{\rm E}[\ell_t(\bm{\theta}_j|X_{s,j},s<t)]\right|\overset{\mathrm{a.s.}}\longrightarrow0,\\
&\sup\limits_{\bm{\theta}_j\in\Theta_j}\left|\frac{1}{n}\widetilde{L}_{n_j}^{'}(0,\bm{\theta}_j)-{\rm E}[\ell_t^{'}(\bm{\theta}_j|X_{s,j},s<t)]\right|\overset{\mathrm{a.s.}}\longrightarrow0,\\
&\sup\limits_{\bm{\theta}_j\in\Theta_j}\left|\frac{1}{n}\widetilde{L}_{n_j}^{''}(0,\bm{\theta}_j)-{\rm E}[\ell_t^{''}(\bm{\theta}_j|X_{s,j},s<t)]\right|\overset{\mathrm{a.s.}}\longrightarrow0.
\end{align*}

$\bullet$ Assumption 4*: For each $j$, $\{\ell_t(\bm{\theta}_j|X_{s,j},s<t);t\in \mathbb{Z}\}$ and $\{\ell_t^{'}(\bm{\theta}_j|X_{s,j},s<t);t\in \mathbb{Z}\}$ are strongly mixing sequences of random variables with geometric rate.

To substantiate the aforementioned assumptions, we then divided the proof into the following three steps.\\
$\textbf{Step 1.}$ We first prove the Assumption 1 (2) to be hold. 
In fact, since BAR(1) model is bounded, we can easily prove that Assumption 1 ($\kappa$) in \cite{DavisYau2013} to be hold for any $\kappa\geqslant1$.
That is, for any $j=1,2,...,m+1$, the function $\ell_t$ is two-time continuously differentiable with respective to $\bm{\theta}_j$, and the first and second derivatives satisfy
\begin{align*}
\sup_{\lambda_d,\lambda_u}\sup\limits_{\bm{\theta}_j\in\Theta_j}|\frac{1}{n}\widetilde{L}_{n_j}(\bm{\theta}_j,\lambda_d,\lambda_u)-\frac{1}{n}L_{n_j}(\bm{\theta}_j,\lambda_d,\lambda_u)|&=o(n^{-1}),\\
\sup_{\lambda_d,\lambda_u}\sup\limits_{\bm{\theta}_j\in\Theta_j}|\frac{1}{n}\widetilde{L}_{n_j}^{'}(\bm{\theta}_j,\lambda_d,\lambda_u)-\frac{1}{n}L_{n_j}^{'}(\bm{\theta}_j,\lambda_d,\lambda_u)|&=o(n^{-1}),\\
\sup_{\lambda_d,\lambda_u}\sup\limits_{\bm{\theta}_j\in\Theta_j}|\frac{1}{n}\widetilde{L}_{n_j}^{''}(\bm{\theta}_j,\lambda_d,\lambda_u)-\frac{1}{n}L_{n_j}^{''}(\bm{\theta}_j,\lambda_d,\lambda_u)|&=o(1),
\end{align*}
almost surely.
\\
$\textbf{Step 2.}$ Assumption 2 (4) and Assumption 3 are the regularity conditions for the conditional log-likelihood
function to  ensure the consistency of the maximum likelihood estimation. Where, similar to the argument in the \textbf{Step 1}, Assumption 2 (4) is obviously true because the BAR(1) model is bounded. Assumption 3 can be verified by the ergodic theorem and the compactness of the parameter space.\\
$\textbf{Step 3.}$ We finally verify Assumption 4* to hold. It is well known that the BAR(1) process is a stationary ergodic Markov chain,
according to the discussion on Page 101 in \cite{Basrak2002}, the BAR(1) process is strongly mixing with geometric rate.
As a result, $\ell_t(\bm{\theta}_j|X_{s,j},s<t)$ is strongly mixing with the same geometric rate as it is a function of finite number
of the strongly mixing $X_{t,j}$'s (Theorem 14.1 of \cite{Davidson1994}). Thus Assumption 4* holds. In fact, Lemma 1 in \cite{DavisYau2013} already states that Assumption 4(0.5) in \cite{DavisYau2013} holds under Assumption 2(2) and 4*. Clearly, Assumption 2(2) is true, so Assumption 4(0.5) in \cite{DavisYau2013} is also satisfied. The proof of Theorem \ref{strong} is completed.

\begin{lemma}\label{lemma1}
Under $\mathcal{H}_0$, we have
$$\max_{1\leqslant k\leqslant n}\sqrt{\frac{k}{n}}\|\Delta_{k}\|=o_{p}(1).$$
\end{lemma}
$\mathbf{Proof~of~Lemma~\ref{lemma1}.}$
Recall that
\begin{align*}
\Delta_{k}=
\begin{cases}	
-\left(\bm{V}_{CLS}-\bm{V}_{k,CLS}\right)\cdot\left(\bm{V}_{k,CLS}\right)^{-1}\cdot\frac{1}{2\sqrt{k}}\frac{\partial S_{k}(\bm{\theta}_0)}{\partial\bm{\theta}},~\text{if}~\bm{V}_{k,CLS}^{-1}~\text{exists},\\
\left(\bm{V}_{CLS}-\bm{V}_{k,CLS}\right)\cdot\sqrt{k}(\hat{\bm{\theta}}_{k,CLS}-\bm{\theta}_0),~\text{otherwise.}
\end{cases}
\end{align*}
Note that the fact $\bm{V}_{n,CLS}\overset{a.s.}{\longrightarrow} \bm{V}_{CLS}$ as $n\rightarrow\infty$ and $\bm{V}_{CLS}$ is a positive definite matrix under $\mathcal{H}_0$.
Therefore, it follows from Egorov's theorem that given $\epsilon>0$ and $\delta>0$, there exists an event $E$ with ${\rm P}(E)>1-\epsilon/3$,
a positive real number $\eta$, and a positive integer $n_0$, such that on $E$ and for all $n>n_0$,
\begin{align}\label{lemma1.1}
v_n\geqslant \eta,
\end{align}
where $v_n$ denotes the minimum eigenvalue of $\bm{V}_{n,CLS}$, and for each $1\leqslant i,j\leqslant 2$
\begin{align}\label{lemma1.2}
\left|\bm{V}_{CLS}^{(i,j)}-\bm{V}_{n,CLS}^{(i,j)}\right|\leqslant \delta\eta.
\end{align}
Since (\ref{lemma1.1}) implies the existence of $\bm{V}_{n,CLS}^{-1}$, we have that on $E$ and for all $n>n_0$,
\begin{align*}
\Delta_{n}=-\left(\bm{V}_{CLS}-\bm{V}_{n,CLS}\right)\cdot\left(\bm{V}_{n,CLS}\right)^{-1}\cdot\frac{1}{2\sqrt{n}}\frac{\partial S_{n}(\bm{\theta}_0)}{\partial\bm{\theta}},
\end{align*}
Thus, on $E$,
\begin{align}\label{lemma1.3}
&\max_{n_0< k\leqslant n}\sqrt{\frac{k}{n}}||\Delta_{k}||
=\max_{n_0< k\leqslant n}\sqrt{\frac{k}{n}}\left\|\left(\bm{V}_{CLS}-\bm{V}_{k,CLS}\right)\cdot\left(\bm{V}_{k,CLS}\right)^{-1}\cdot\frac{1}{2\sqrt{k}}\frac{\partial S_{k}(\bm{\theta}_0)}{\partial\bm{\theta}}\right\|\nonumber\\
&\leqslant \max_{n_0< k\leqslant n}\left\|\bm{V}_{CLS}-\bm{V}_{k,CLS}\right\|\cdot
\max_{n_0< k\leqslant n}\left\|\left(\bm{V}_{k,CLS}\right)^{-1}\right\|\cdot
\max_{n_0< k\leqslant n}\left\|\frac{1}{2\sqrt{n}}\frac{\partial S_{k}(\bm{\theta}_0)}{\partial\bm{\theta}}\right\|\nonumber\\
&\leqslant \sum_{i=1}^2
\max_{n_0< k\leqslant n}\left\|\bm{V}_{CLS}-\bm{V}_{k,CLS}\right\|\cdot
\max_{n_0< k\leqslant n}\left\|\left(\bm{V}_{k,CLS}\right)^{-1}\right\|\cdot
\max_{1\leqslant k\leqslant n}\left|\frac{1}{2\sqrt{n}}\frac{\partial S_{k}(\bm{\theta}_0)}{\partial\bm{\theta}_i}\right|.
\end{align}
First, from (\ref{lemma1.2}) it holds that on $E$ and for $n>n_0$,
\begin{align*}
\left\|\bm{V}_{CLS}-\bm{V}_{n,CLS}\right\|&:=\sup\left\{\left\|(\bm{V}_{CLS}-\bm{V}_{n,CLS})h\right\|:\|h\|\leqslant 1\right\}\\
&=\sup\left\{\left(\sum_{i=1}^2\left(\sum_{j=1}^2(\bm{V}_{CLS}^{(i,j)}-\bm{V}_{n,CLS}^{(i,j)})h_j\right)^2\right)^{1/2}:
\sum_{j=1}^2h_j^2\leqslant 1\right\}\\
&\leqslant 2\delta\eta
\end{align*}
and consequently, on $E$
\begin{align}\label{lemma1.4}
\max_{n_0< k\leqslant n}\left\|\bm{V}_{CLS}-\bm{V}_{k,CLS}\right\|\leqslant 2\delta\eta.
\end{align}
Next, since $\bm{V}_{n,CLS}$ is a real symmetric matrix for all $n\geqslant 1$ and (\ref{lemma1.1}) holds on $E$ and for all $n>n_0$, we have
\begin{align}\label{lemma1.5}
\max_{n_0< k\leqslant n}\left\|\left(\bm{V}_{k,CLS}\right)^{-1}\right\|
&=\max_{n_0< k\leqslant n} \left\{\text{the maximum absolute eigenvalue of} \left(\bm{V}_{k,CLS}\right)^{-1}\right\}\nonumber\\
&=\max_{n_0< k\leqslant n} \frac{1}{v_n}\leqslant \frac{1}{\eta}.
\end{align}
Therefore, combing equations (\ref{lemma1.4}) and (\ref{lemma1.5}), it follows that the right-hand side of (\ref{lemma1.3}) is no more than
$\sum_{i=1}^2 2\delta\max_{1\leqslant k\leqslant n}\left|(2\sqrt{n})^{-1}\partial S_{k}(\bm{\theta}_0)/\partial\bm{\theta}_i\right|$
on $E$, and consequently,
\begin{align*}
{\rm P}\left\{\left(\max_{n_0< k\leqslant n}\sqrt{\frac{k}{n}}\|\Delta_{k}\|>\frac{\epsilon}{2}\right)\bigcap E\right\}
&\leqslant{\rm P}\left\{\sum_{i=1}^2 2\delta\max_{1\leqslant k\leqslant n}\left|\frac{1}{2\sqrt{n}}\frac{\partial S_{k}(\bm{\theta}_0)}{\partial\bm{\theta}_i}\right|>\frac{\epsilon}{2}\right\}\\
&\leqslant\sum_{i=1}^2{\rm P}\left\{\max_{1\leqslant k\leqslant n}\left|\frac{\partial S_{k}(\bm{\theta}_0)}{\partial\bm{\theta}_i}\right|>\frac{\epsilon\sqrt{n}}{4\delta}\right\}\\
&\leqslant\sum_{i=1}^2\frac{16\delta^2}{n\epsilon^2}{\rm E}\left(\frac{\partial S_{n}(\bm{\theta}_0)}{\partial\bm{\theta}_i}\right)^2.
\end{align*}
Clearly, under $\mathcal{H}_0$, there exist a constant $0<c<\infty$, such that
$$\sum_{i=1}^2\frac{1}{n}{\rm E}\left(\frac{\partial S_{n}(\bm{\theta}_0)}{\partial\bm{\theta}_i}\right)^2\leqslant c.$$
Furthermore, we set $\delta\leqslant\sqrt{\epsilon^3/48c}$, and there is
\begin{align*}
{\rm P}\left\{\left(\max_{n_0< k\leqslant n}\sqrt{\frac{k}{n}}\|\Delta_{k}\|>\frac{\epsilon}{2}\right)\bigcap E\right\}
\leqslant\frac{\epsilon}{3}.
\end{align*}
Notably, there exists a positive integer $n_1>n_0$, such that
\begin{align*}
{\rm P}\left\{\max_{1\leqslant k\leqslant n_0}\sqrt{\frac{k}{n_1}}\|\Delta_{k}\|>\frac{\epsilon}{2}\right\}
\leqslant\frac{\epsilon}{3}.
\end{align*}
Following the above derivation, we have that for all $n>n_1$
\begin{align*}
{\rm P}\left\{\max_{1\leqslant k\leqslant n}\sqrt{\frac{k}{n}}\|\Delta_{k}\|>\epsilon\right\}
\leqslant&{\rm P}\left\{\max_{1\leqslant k\leqslant n_0}\sqrt{\frac{k}{n}}\|\Delta_{k}\|+\max_{n_0< k\leqslant n}\sqrt{\frac{k}{n}}\|\Delta_{k}\|>\epsilon\right\}\\
\leqslant& {\rm P}\left\{\max_{1\leqslant k\leqslant n_0}\sqrt{\frac{k}{n}}\|\Delta_{k}\|>\frac{\epsilon}{2}\right\}\\
&+{\rm P}\left\{\left(\max_{n_0< k\leqslant n}\sqrt{\frac{k}{n}}\|\Delta_{k}\|>\frac{\epsilon}{2}\right)\bigcap E\right\}+{\rm P}\{E^c\}
\leqslant \epsilon,
\end{align*}
which implies
$$\max_{1\leqslant k\leqslant n}\sqrt{\frac{k}{n}}\|\Delta_{k}\|=o_{p}(1).$$
The proof of Lemma \ref{lemma1} is completed.
~\\
~\\
\underline{\textbf{Searching algorithm based on genetic algorithm}}
\begin{table}[H]
		\begin{tabular}{ll}
			\toprule
			\multicolumn{2}{l}{$\textbf{S step:}$}\\
			\cmidrule(lr){1-2}
\textbf{Input}:& The random sample $\{x_1, x_2, . . . , x_n\}$ from the MCP-BAR(1) model.\\
&The upper bound $N$ in MCP-BAR(1) model.\\
&The upper bound of the number of change-points $M_0$.\\
&The sample size $n$.\\
\textbf{Initialise}:& Let $^\ast$$\textbf{MDL-best}_1=\min\limits_{\bm{\tau}, \bm{\theta}}\textbf{MDL}(1,\frac{\bm{\tau}}{n},\bm{\theta})$, where \textbf{MDL} is defined by (\ref{MDL});\\
&Let $\hat{m}=1$, and $(\hat{\bm{\tau}},\hat{\bm{\theta}})=\arg \min\limits_{\bm{\tau}, \bm{\theta}}\textbf{MDL}(1,\frac{\bm{\tau}}{n},\bm{\theta})$.\\
\textbf{Iterate}:& for $j=2:M_0$\\
&~~~ Let $^\ast$$\textbf{MDL-best}_2=\min\limits_{\bm{\tau},\bm{\theta}}\textbf{MDL}(j,\frac{\bm{\tau}}{n},\bm{\theta})$.\\
&~~~ Let $(\hat{\bm{\tau}}^{\star},\hat{\bm{\theta}}^{\star})=\arg\min\limits_{\bm{\tau},\bm{\theta}}\textbf{MDL}(j,\frac{\bm{\tau}}{n},\bm{\theta})$.\\
&~~~~~~ if $\textbf{MDL-best}_2 > \textbf{MDL-best}_1$\\
&~~~~~~~~~~break.\\
&~~~~~~else\\
&~~~~~~~~~~$\textbf{MDL-best}_1 =\textbf{MDL-best}_2$.\\
&~~~~~~~~~~$\hat{m}=j$, and $(\hat{\bm{\tau}},\hat{\bm{\theta}})=(\hat{\bm{\tau}}^{\star},\hat{\bm{\theta}}^{\star})$.\\
&~~~~~~ end\\
&end\\
\textbf{Output}:& change-points estimator $(\hat{m},\hat{\bm{\tau}},\hat{\bm{\theta}})$.\\
\toprule
\end{tabular}
\begin{tabular}{l}
\toprule
$^\ast\textbf{GA step:}$\\
\cmidrule(lr){1-1}
\multirow{2}{15.4cm}{\textbf{Step 1}: If $\bm{\tau}$ is given, the estimator for the $j$th segment parameter $\hat{\bm{\theta}}_j$  can be easily obtained by the closed-form expressions or solved by MATLAB function ``\textbf{fmincon}".}\\
~\\
\multirow{1}{15.4cm}{\textbf{Step 2}: Use ``ga" function in MATLAB to get the global optimal solution $\hat{\bm{\tau}}$.}\\
\toprule
\multirow{3}{15.4cm}{$^\ast$When $(m, \bm{\tau})$ is fixed, parameter estimation for each segment of our model is a conventional optimization problem. We divided the estimation procedure into the above two steps.}\\
~\\
~\\
\end{tabular}
\end{table}
$\textbf{Setting tips for \textbf{Step 1} and \textbf{Step 2}:}$
\begin{enumerate}[($1$)]
\setlength{\itemsep}{0pt}
\setlength{\parskip}{0pt} 
\setlength{\parsep}{0pt} 
\item Since the closed-form of the estimators will greatly improve the computation speed, we minimized MDL based on the CML likelihood function evaluated at CLS estimators (\ref{cls_closed-form}) in \textbf{Step 1}.
\item The options of ``ga" function is set as:
``options$=$gaoptimset(`PlotFcns',\{@gaplotbestf\},`Popula\\tionSize',10*$s$, `CrossoverFraction', 0.55, `Generations', 300)",
where `PopulationSize' is the size of the population, `CrossoverFraction' (CF) is the fraction of the population at the next generation, not including elite children, that the crossover function creates. Later in Section 5, we will do the  sensitivity analysis of the tunning parameter CF.
\item To satisfy the \textbf{Assumption \ref{assumption1}}: each segment must have a sufficient number of observations, the following restriction is given when we carry out \textbf{Step 2}:\\
if $\min(\hat{\tau}_{i+1}-\hat{\tau}_i)<n*\epsilon_{\lambda}, (i=1,...,j)$, we set \textbf{MDL}$(\cdot,(\hat{\tau}_1,...,\hat{\tau}_j),\cdot)=\text{INF}$. In simulation, we set $\epsilon_{\lambda}=10/n.$
\end{enumerate}

\end{document}